\documentclass[letterpaper,fleqn,usenatbib]{mnras}

\usepackage{enumerate}
\usepackage{graphicx}	
\usepackage{amsmath}	

\def\adhoc{{\it ad hoc}}
\def\eg{{\it e.g.}}
\def\etal{{\it et al.}}
\def\etc{{\it etc.}}
\def\ie{{\it i.e.}}

\def\DFt{{\tt DiskFit}}
\def\mkm{{\tt makemap}}
\def\rtc{{\tt rotcur}}
\def\smk{{\tt setmask}}

\def\pmb#1{\setbox0=\hbox{$#1$}%
  \kern-0.25em\copy0\kern-\wd0
  \kern.05em\copy0\kern-\wd0
  \kern-0.025em\raise.0433em\box0}
\def\spmb#1{\setbox1=\hbox{${\scriptstyle #1}$}%
  \kern-0.25em\copy1\kern-\wd1
  \kern.05em\copy1\kern-\wd1
  \kern-0.025em\raise.0433em\box1}

\title[Galaxy Rotation Curves]{Uncertainties in Galaxy Rotation Curves}

\author[Sellwood, Spekkens \& Eckel]{
J. A. Sellwood,$^{1}$\thanks{E-mail:sellwood@as.arizona.edu}
Kristine Spekkens,$^{2}$\thanks{E-mail:kristine.spekkens@gmail.com}
and
Carter S. Eckel,$^1$\thanks{E-mail:cartereckel@email.arizona.edu}
\\
$^1$Steward Observatory, University of Arizona, 933 N Cherry Ave,
Tucson AZ 85722, USA  \\ $^2$Department of Physics and Space Science, Royal Military College of Canada, Box 17000, Station Forces, Kingston, ON K7K 7B4, Canada}

\pubyear{2020}

\begin{document}
\label{firstpage}
\pagerange{\pageref{firstpage}--\pageref{lastpage}}
\maketitle

\begin{abstract}
Assessing the likelihood that the rotation curve of a galaxy matches
predictions from galaxy formation simulations requires that the
uncertainties in the circular speed as a function of radius derived
from the observational data be statistically robust.  Few
uncertainties presented in the literature meet this requirement.  In
this paper we present a new standalone tool, \mkm, that estimates the
fitted velocity at each pixel from Gauss-Hermite fits to a 3D spectral
data cube, together with its uncertainty obtained from a modified
bootstrap procedure.  We apply this new tool to neutral hydrogen
spectra for 18 galaxies from the THINGS sample, and present new
velocity maps with uncertainties.  We propagate the estimated
uncertainties in the velocity map into our previously-described model
fitting tool \DFt\ to derive new rotation curves.  The uncertainties
we obtain from these fits take into account not only the observational
errors, but also uncertainties in the fitted systemic velocity,
position of the rotation centre, inclination of the galaxy to the line
of sight, and forced non-circular motion.  They are therefore much
better-defined than values that have previously been available.  Our
estimated uncertainties on the circular speeds differ from previous
estimates by factors ranging up to of five, being smaller in some
cases and larger in others.  We conclude that kinematic models of
well-resolved HI datasets vary widely in their precision and
reliability, and therefore potentially in their value for comparisons
with predictions from cosmological galaxy formation simulations.
\end{abstract}



\section{Introduction}
\label{sec.intro}
Galaxy formation simulations \citep[see][for a recent review]{SD15}
are making increasingly detailed predictions for the properties of
galaxies to be confronted with data.  Much of the recent emphasis has
been on chemical evolution and the distribution of metals among the
stars and both the cool and hot gas, and plenty of data is coming from
surveys such as APOGEE \citep{Ma17} and GALAH \citep{deS15}.  However,
the structure and mass distribution of disc galaxies remain key
predictions of the models \citep[\eg][]{Buck20, Vogel20, Well20} that
can be confronted by spectral line observations of galaxies in the
local universe.  The 21cm line of atomic hydrogen (HI) is particularly
useful because the HI disc is typically more extended than the optical
galaxy and thus traces the kinematics into the region dominated by
dark matter.  As a result, there is now a long history of detailed
comparisons between the observed rotation curves of nearby galaxies
derived from interferometric HI maps and theoretical predictions from
galaxy formation models \citep[see][for reviews]{deB10, Le16}.

Much recent discussion has focused on whether or not baryonic effects,
such as feedback from star formation, can explain discrepancies
between the predicted mass distributions of collisionless cold dark
matter halos and the observed rotation curves of nearby galaxies.  The
slope and scatter of scaling relations built from rotation curve
compilations are powerful tools in this regard \citep{Du07, Le16,
  Po19}, and it remains unclear whether rotation curve models of
individual galaxies are, or are not, consistent with cosmological
galaxy formation predictions \citep[\eg][]{Oh11, Oman15, Ka17, Ha18,
  Ma20}.  Some issues raised may be somewhat less affected by baryonic
feedback, which is frequently invoked to explain away the cusp/core
controversy and other small-scale problems \citep[reviewed
  by][]{We15}.

Properly evaluated uncertainties in the measured circular speeds are
required in order to assess the likelihood that a set of observed
rotation curves from galaxies is consistent with the predictions of
the models.  However, such uncertainties have not as yet been derived
in a statistically correct manner.

The traditional procedure for deriving a rotation curve from a
well-resolved HI data cube of a nearby, intermediate inclination
galaxy has been first to create a 2D velocity map and then to fit a
tilted ring model to the velocity map \citep{RLW}.  The standard
tilted ring software, \rtc\ \citep{Be89}, assumes the gas to flow on
circular orbits.  \citet{FvGdZ} and \citet{Sc97} generalized \rtc\ to
mildly non-circular flows, but their tool, {\tt reswri}, assumes that
the response to potential distortions can be modelled by forced
epicycles, and can therefore fit only mildly elliptical streamlines.

The underlying assumption of such models is that the gas is flowing in
a thin, possibly warped, layer in centrifugal balance in a nearly
axisymmetric potential, and therefore has a well-defined circular
speed at every radius in the galaxy disc.  Even accepting that these
assumptions are true, the fitted circular speed at each radius is
affected by the choices for the systemic velocity, the location of the
rotation centre, and both the inclination and position angle of the
disc plane to the line of sight.  Uncertainties in these quantities,
as well as beam smearing, turbulence, forced non-axisymmetric flows,
and noise in the data, should all contribute to the uncertainty in the
estimated circular speed at each radius.

As the resolution and S/N of the observations have improved, the
formal statistical uncertainties in both the estimated velocity in
each pixel of the 2D map and the fitted circular speed at each radius
have become absurdly small, for the reason we give below in
\S\ref{sec.boots}.  Because of this, many workers
\citep[\eg][]{Ch09} present {\it ad hoc} error bars on the circular
speed estimates at each radius in the galaxy that are based upon the
difference between the separately-fitted circular speeds on the
approaching and receding sides of the galaxy, while holding the
kinematic centre, systemic velocity, and projection angles at their
best fit values for the whole map. \citet{deB08} adopt this approach
but add in quadrature the dispersion of velocities within the model
ring.  Although these ``error bars'' have little statistical validity,
they are claimed to be realistic merely because the values are a
significant fraction of the fitted circular speed and are generally
larger where the model is a poorer fit.

Several algorithms are now publicly-available that analyze the 3D
spectral cube directly rather than 2D velocity fields \citep{Kamp15,
  DiT15, Davi13, Davi20}, building on earlier development by
\citet{Jozsa07}.  This approach is particularly useful for comparing
complex models to deep, very well-resolved HI data \citep{Jozsa09,
  Kamp13, Ma19} or for lower resolution data in which beam smearing in
2D maps can be severe \citep{DiT15, Kamp15}. In principle, 3D models
constrain HI spectral data more directly than their 2D counterparts
and avoid the need to first derive a 2D velocity field.  But in
practice most apply the same iterative approach of the classic 2D
\rtc/{\tt reswri} algorithm \citep{Si97, Sc97} as has long been done
in 2D, and suffer from the same shortcomings as described above.
Uncertainties returned by frequentist codes generally stem from
\adhoc\ variations of model parameters \citep[\eg][]{Kamp15, DiT15},
and there is a need to explore whether the posterior distributions of
Bayesian methods \citep[\eg][]{Bouche15, Davi17} better reflect
uncertainties in the fit.  In addition, performance tests on
well-resolved datasets have revealed that 3D codes converge much more
slowly to the same fitting parameters as are found from 2D velocity
maps.  3D methods are therefore impractical in this regime and there
remains a need for statistically robust techniques to derive and
analyze 2D velocity maps from 3D HI spectral cubes.

As already noted, fitting an idealized circular flow pattern to a 2D
kinematic map neglects the possible existence of coherent turbulence
or noncircular motions forced by spiral arms and the like
\citep{Oman19}.  Thus the residuals after subtracting a circular flow
model contain correlations that cannot be accounted for in the formal
statistical errors of the model parameters.  The advantages of the
\DFt\ software \citep{SS15} over \rtc\ are two fold: (1) it can fit
for bar-like flows and (2) it also estimates uncertainties in all
fitted parameters and the fitted velocities at each radius.  The
modified bootstrap procedures described by \citet{SS07} and
\citet{SZS10} to estimate all uncertainties not only take account of
the large-scale correlations in the pattern of residuals after
subtracting a fitted smooth model from the velocity map but also
factor in uncertainties in the global parameters.  It should be noted
that the uncertainties yielded by their procedures seemed reasonable,
but these authors did not attempt to prove that they were large enough
or statistically meaningful.

However, uncertainties in the fitted velocity at each pixel of the
velocity map are generally unavailable for neutral hydrogen data, and
therefore cannot be propagated into the uncertainties in the rotation
curve.  When this is the case, \DFt\ adopts a fixed uncertainty,
$\Delta_{\rm ISM}$, whose value is chosen by the user.  Values in the
range $8 \la \Delta_{\rm ISM} \la 12\;$km~s$^{-1}$ would be consistent
with the observed turbulent motions of gas in the ISM of a large star
forming disc galaxy \citep{GKT79, Ka93}.  This choice is motivated by
the reasoning that the velocity fitted to the spectrum may be
dominated by emission from material whose velocity may differ from the
circular speed by this much.  However, an assumption of a uniform
uncertainty in the fitted velocity of each pixel is clearly
inadequate, since it gives equal weight in the fitted model to pixels
at which the velocity is well-determined and to those where the fitted
velocity is poorly constrained.

While bootstrap iterations in \DFt\ yield uncertainties in the fitted
parameters and velocities, it would be better to begin with a velocity
map that also has an associated uncertainty for the line-of-sight
velocity at each pixel.  Once velocity errors in each pixel of the map
are in hand, they can be propagated all through the fitting process to
yield more statistically meaningful uncertainties in the derived
projection parameters and circular speed at each radius, as
we describe in \S\ref{sec.DiskFit}.

In \S\ref{sec.makemap}, we describe a new standalone program, \mkm,
that derives a 2D velocity map from a 3D data cube.  It uses a
modified bootstrap method to estimate the uncertainties in the
velocity at each pixel.  We distribute it as an addition to the
\DFt\ package \citep{SS15}.  There seems no reason, in principle, why
the modification to the bootstrap method we develop here could not be
employed in other packages that attempt to extract the rotation curve
directly from the data cube.

We apply \mkm\ to 18 publicly available data cubes of HI emission
from galaxies in the THINGS survey \citep{Wa08}.  From each data cube,
we extract a 2D velocity map together with a map of the uncertainty in
the estimated velocity at each pixel, and input the derived maps to
\DFt\ to obtain new rotation curves having uncertainties that are
better-determined.

\section{Making a map}
\label{sec.makemap}
This section introduces program \mkm, which fits a smooth
Gauss-Hermite function to the line profile in each velocity column of
the data cube that appears to contain significant spectral line
emission.  We use a modified bootstrap approach to estimate
uncertainties for the fitted line parameters.  In this section we
illustrate our procedure using the THINGS data cube for NGC~2841;
we apply \mkm\ to 18 THINGS galaxies in \S\ref{sec.results}.

A number of different methods have been employed to derive a 2D velocity map
from a 3D data cube, which also differ between optical and radio data.
On the one hand, the spectral resolution of HI emission line
data is high enough that individual velocity channels are
statistically independent, something that is not usually true for
optical data from a Fabry-P\'erot system or a fibre bundle, say.  On
the other hand, the spatial resolution of optical data is generally
high enough that adjacent pixels are independent, or nearly so, while
the beam size in radio data generally extends over a number of pixels.
Here we focus on HI data from interferometers, and leave
applications of the ideas we develop to CO data or to optical data to
later papers.

The intensity-weighted mean velocity, also known as the first moment,
is one possible estimate of the gas velocity, and the THINGS team
\citep{Wa08} have posted moment 1 maps of all the galaxies in that
sample on their website.  Some workers \citep[\eg][]{Ge04} trace the
envelope of the line profile to find the velocity of peak emission.
Others \citep[\eg][]{Po16} fit a single Gaussian to the line profile,
adopting the peak of the fitted function as the derived velocity.
\citet{Oh08, Oh19} fit multiple Gaussians to try to separate side
peaks, which they attribute to random motions, from the ``bulk''
velocity.  \citet[][hereafter dB08]{deB08} fitted a generalized
Gaussian \citep{Ge93, vdMF93}, which has the advantage that it takes
account of the possible skewness of the line profile that can be
significant, especially for large beam sizes and when the galaxy is
not far from edge-on.  dB08 presented two examples to illustrate the
differences between the velocity estimated by their new method and
those obtained by the other procedures.  Since the formal statistical
uncertainty in the fitted velocity at each pixel in the map is always
much smaller than the likely turbulent dispersion, many authors
neglect velocity uncertainties altogether.

\subsection{Line fitting}
Following dB08, we fit a Gauss-Hermite function to the measured
intensity as a function of velocity for each pixel in the data cube.
Assuming only $h_3$ and $h_4$ are non-zero, \citet[][ their
  eq.~(9)]{vdMF93} write the ``line profile'' as
\begin{equation}
{\cal L}(v) = {\cal L}_0 + \left[ {A\alpha(w) \over \sigma} \right] \left\{ 1 + h_3H_3(w) + h_4H_4(w) \right\},
\label{eq.profile}
\end{equation} 
where $w= (v-\bar V) / \sigma$, $H_3(w) = w(2w^2-3)/\sqrt{3}$, $H_4(w)
= (4w^4 - 12w^2 + 3)/\sqrt{24}$, and $\alpha(w) =
\exp(-w^2/2)/\sqrt{2\pi}$.  The set of paramaters, $\{p(m)\}$, to be
determined is: $p(1) = {\cal L}_0$ the estimated continuum, $p(2) = A$
the normalization of the line amplitude, $p(3) = \bar V$ the central
velocity of the fitted profile, $p(4) = \sigma$ an estimate of the
line width, and, optionally, $p(5) = h_3$ and $p(6) = h_4$ the
coefficients that quantify departures of the line profile from a
simple Gaussian.  A skew line profile would be better fitted with a
non-zero value of $h_3$, while $h_4$ would be positive if the profile
has heavy tails and negative for a line that is more peaked than a
Gaussian; these parameters therefore describe the profile in an
analagous, but distinct, manner to the familiar skew and kurtosis
parameters \citep[\eg][]{Press92}.

It is important to realize that the velocity $V_m$ for which
expression (\ref{eq.profile}) is maximum is generally {\it not}\/ the
value of $\bar V$ when $h_3 \neq 0$.  We find that $|V_m - \bar V| \la
\sigma/2$ when $|h_3|\sim 0.5$.  We adopt $V_m$ as ``the best fit
velocity''.  Obviously, $V_m \equiv \bar V$ for a Gaussian function.

Program \mkm\ attempts to fit expression (\ref{eq.profile}) to the set
of intensities $\{I(v_k)\}$ in each spectral column of a data cube,
where $\{v_k\}$ are the channel velocities that are separated by a
fixed $\delta v$, aka the channel width.  We generally recommend that
the fit includes the $h_3$ term only, but the user can choose to fit
for both $h_3$ and $h_4$, or neither (\ie\ a simple Gaussian) if
desired.

The first step is to estimate the noise level, ${\cal N}$, in the raw
data cube, which we assume to be constant.  Program \mkm\ uses the
bi-weight estimator \citep{BFG90} to compute the mean and dispersion
of the intensity values in the outermost 2 layers of all 6 faces of
the data cube, and adopts ${\cal N}$ for the entire cube to be this
estimated dispersion. The bi-weight is superior to the rms value since
it ignores outliers and heavy tails in the distribution of values; see
\citet{BFG90} for a discussion of different distribution width
estimators.

Then working over the image plane, program \mkm\ attempts to fit a
line profile to $\{I(v_k)\}$ only when the maximum value of $I(v_k) >
I_{n, \rm min} \times\, {\cal N}$, with $I_{n, \rm min}$ being
a constant set by the user.

\begin{figure}
\includegraphics[width=\hsize,angle=0]{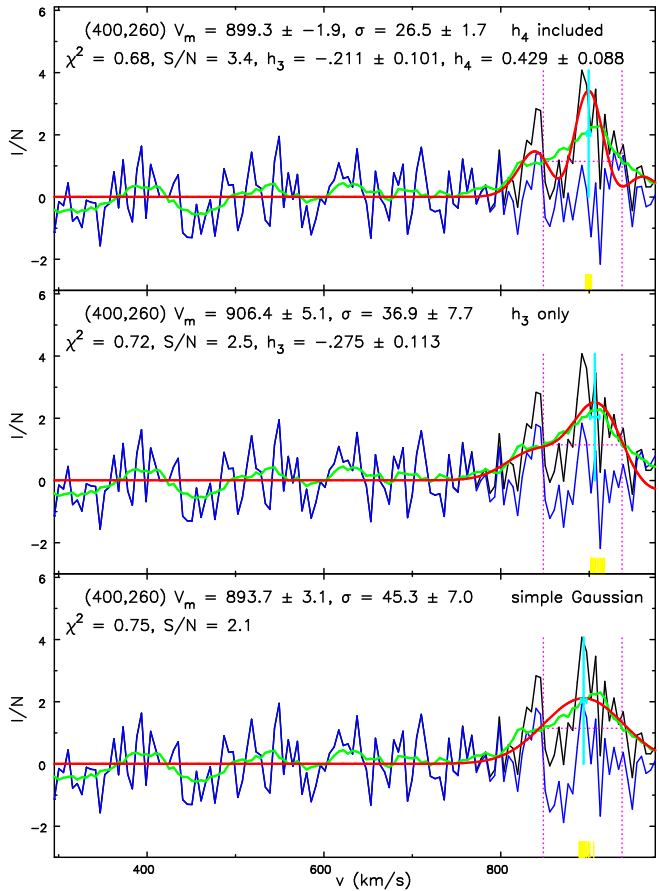}
\caption{An example of a low S/N line from NGC 2841 to illustrate the
  behaviour of the fitting program.  The bottom panel fits a simple
  Gaussian ($h_3=h_4=0$) to the data; that in the middle includes
  $h_3$ only, while the top panel shows the result of fitting for both
  $h_3$ and $h_4$.  In all three panels, the black jagged line shows
  the input data $\{I(v_k)/{\cal N}\}$ in units of the noise, the blue
  jagged line, which obscures the black line for much of the range,
  indicates the residuals after subtracting the fitted function, shown
  by the smooth red curve.  The green line is the 10-point running
  average of $\{\bar I_n(v_k)\}$ and the vertical parts of the dotted
  magenta H illustrate the first guesses at $\bar V \pm \sigma$.
  Finally, the vertical cyan line is the fitted $V_m$ and the
  horizontal error bar shows its estimated uncertainty, while the
  yellow marks indicate fitted maximum velocity from each of the 20
  bootstraps.}
\label{fig.linefit1}
\end{figure}

Program \mkm\ uses the fitting tool {\tt sumsl} \citep[from the
  software collection made public by][]{Bu17}, which is a quasi-Newton
method requiring first derivatives, to find the set of parameters
$\{p(m)\}$ that mimimize
\begin{equation}
\chi_{\rm line}^2 = \sum_k \left[ I_n(v_k) - {\cal L}(v_k) \right]^2,
\label{eq.chisql}
\end{equation}
where now $I_n(v_k) = I(v_k)/{\cal N}$; because the input data values
have been normalized by the noise, this is the usual definition of
$\chi^2$.  This form is appropriate only when the noise level is
constant throughout the data cube, as it is for most HI data cubes. To
use \mkm\ for cubes having a position and/or channel dependent noise
level, one would need to rewrite eq.~(\ref{eq.chisql}) in standard
form.

Starting the minimization from a good first guess helps to prevent the
minimizer from finding a local minimum far from the global minimum.
In order to choose this guess, \mkm\ first creates a smoothed set of
intensities $\bar I_n(v_k)$ from a running average of ten $I_n(v_k)$
values and sets the first guesses of $A$ and $\bar V$ respectively
from the maximum of $\bar I_n(v_k)$ and its velocity.  The initial
guess for $\sigma$ is half the velocity-width of $\bar I_n(v_k)$ at
half its maximum intensity.  The initial guesses for ${\cal L}_0$,
$h_3$, and $h_4$ are all zero.  Although this procedure would have
difficulty in fitting an extremely broad, low S/N line, the sources of
such lines are unlikely to be found in galaxy discs.  For other
applications, we would recommend combining velocity channels before
fitting such lines in order to improve S/N.

Figure~\ref{fig.linefit1} illustrates the fit for one pixel having low
S/N emission in the THINGS data cube for NGC~2841.  The bottom panel
fits a simple Gaussian ($h_3=h_4=0$) to the data; that in the middle
includes $h_3$ only, while the top panel shows the result of fitting
for both $h_3$ and $h_4$.  The caption describes the various coloured
lines.  The fitted parameter values and their uncertainties are
recorded in each panel; note that the negative uncertainty in the
fitted velocity in the top panel flags the bimodal nature of the
fitted function (see \S\ref{sec.bimod}).

Notice that the best fit value of $V_m$ differs by almost 13~km/s
between the Gaussian fit and that which includes $h_3$, which is just
about consistent with the combined estimated uncertainties of the two
fits.  We determine uncertainties as described in \S\ref{sec.boots}.

The user must choose values for a few parameters to determine whether
a line is rejected.  They are: $I_{n, \rm min}$, the minimum
normalized intensity in the spectrum before a fit is attempted,
$A_{\rm min}$, the minimum acceptable line amplitude, $w_{\rm max}$,
the maximum allowed line width, and $h_{\rm max}$, the max acceptable
absolute value of the Gauss-Hermite coefficients.  Tallies of rejected
fits are tabulated according to the following fail criteria:
{\parskip=0pt
\begin{enumerate}[1]
\item The maximum value of $I_n(v_k) < I_{n, \rm min}$,
\item The line fitting routine fails (very rare),
\item The line intensity, in units of ${\cal N}$,
  $A/(\sqrt{2\pi}\sigma) < A_{\rm min}$,
\item The fitted $\bar V$ is outside the range of velocities in the
  channel maps,
\item The line width $\sigma > w_{\rm max}\, \delta v$.  The fit is
  also rejected when $\sigma < 2I_{n, \rm min}\,\delta v / I_{n, \rm
    max}$, where $I_{n, \rm max}$ is the greatest value of
  $\{I_n(v_k)\}$.  This is to reject low S/N lines, that could be
  little more than a noise spike in one channel, while retaining
  high S/N lines that have a narrow width.
\item Either $|h_3|> h_{\rm max}$ or $|h_4| > h_{\rm max}$,
\item When bootstraps are included (see \S\ref{sec.boots} below), one
  or more iterations has returned nonsensical fitted parameters.  To
  wit, either $A<0$, $\sigma<0$, or $\bar V$ is outside the range of
  velocities in the channel maps.  This usually happens only when the
  S/N is low.
\end{enumerate}}

\noindent Program \mkm\ reports these tallies, as well as the number
of line profiles that were successfully fitted.  For every fit that
fails, it assigns the default values of zero for all parameters.

We determine the best fit velocity, $V_m$, from among the roots of
$\partial {\cal L} / \partial v = 0$.  Rejected fits are assigned the
default value $V_m = -9999\;$km/s.

\begin{figure}
\includegraphics[width=\hsize,angle=0]{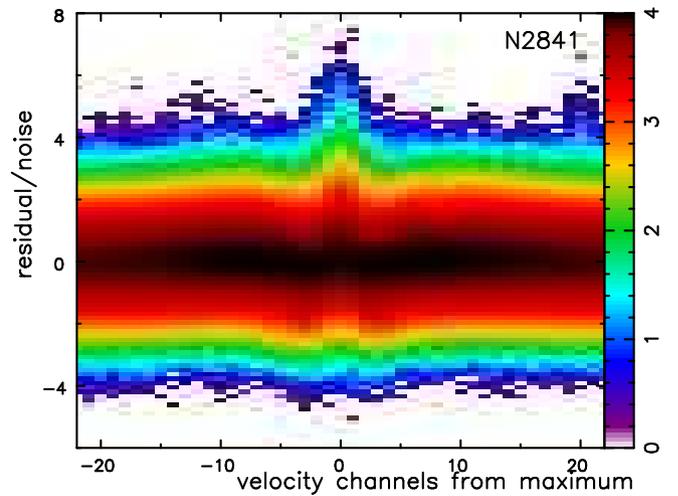}
\caption{The distributions of residuals from all fitted pixels in the
  data cube from NGC~2841, in units of the noise, after subtracting
  the best fit function (\ref{eq.profile}) that included only the
  $h_3$ term. The distribution of residuals in the velocity channels
  at each pixel is shifted so that the channel with the greatest
  fitted intensity is at 0. The colour represents the logarithm of the
  number of pixels having the given value.  The distributions of
  residuals are consistent with noise except for a few channels near
  the line centre.}
\label{fig.errdist}
\end{figure}

\subsection{Parameter uncertainties}
\label{sec.boots}
Program \mkm\ estimates the uncertainty in each fitted parameter by a
modified bootstrap method.  The user must supply a value for $n_{\rm
  boot}$, the number of desired bootstrap iterations to be used for
each pixel of the map for which the fit was accepted.  A value of
$n_{\rm boot}\leq0$ will prevent uncertainties from being estimated.

As noted in \S\ref{sec.intro}, the principal obstacle to a
satisfactory accounting for errors is that the formal statistical
errors in both the fitted velocity at each pixel and the circular
speed at each radius are absurdly small.  It is instructive to
consider why this should be the case.

When fitting a parameterized model to data by mimimizing $\chi^2$,
MCMC, or any other method, the standard estimators of the parameter
uncertainties assume that the data contain no additional information
other than the model plus noise or, in other words, $\chi^2$ per
degree of freedom should be close to unity.  When this ideal
assumption holds, errors estimated by formal techniques should be
meaningful.  The assumption that the data are perfectly described by
the adopted model plus uncorrelated noise rarely holds in the real
world, however, where the model is generally an idealized
approximation to the data.

In the case of intensities as a function of velocity in each pixel
spectrum of a data cube, the line profile is indeed noisy because of
imperfections in the observations, but the net emission from a clumpy,
turbulent ISM is most unlikely to have the kind of intrinsically
smooth profile that is fitted, and emission from individual clouds
must add to the ragged appearance of the data, especially at the
velocities close to the line peak.  Subtracting a smooth line profile
from the data and considering the residuals to be exclusively noise,
neglects the intrinsic substructure in the line profile, as was
already recognized by \citet{Oh08, Oh19}; the true uncertainty in the
fitted velocity is increased when substructure is accounted for.

The standard bootstrap strategy to estimate uncertainties is to
subtract the fitted function, eq.~(\ref{eq.profile}), from the data to
create a set of residuals $\{\delta(v_k) \equiv I_n(v_k) - {\cal
  L}(v_k)\}$ at each $v_k$, and then make new fits to $n_{\rm boot}$
sets of pseudo data $P(v_j)$, where $P(v_j) = {\cal L}(v_j) +
\delta^\prime(v_j)$ to find $n_{\rm boot}$ sets of parameters
$\{p^\prime(m)\}$ and $V_m^\prime$.  At each iteration
$\delta^\prime(v_j)$ is a random draw from the pool of
$\{\delta(v_k)\}$.  The uncertainty estimate of each parameter is then
$ep(m) = \langle[p(m)-p^\prime(m)]^2\rangle^{1/2}$ and similarly for
$eV_m$.

This procedure yields the correct uncertainty when the observed line
is perfectly described by the adopted function (eq.~\ref{eq.profile}),
and the residuals are simply uncorrelated noise.  However, the pattern
of residuals around the line peak contains some signal that is more
than just random noise, as is illustrated in Figure~\ref{fig.errdist}.
This figure is based upon all 163,676 pixels for which $h_3$ only fits
to the THINGS data cube from NGC~2841 were accepted.  For each pixel,
the distribution of residuals, divided by the noise, is shifted in
velocity so that the peak of the fitted line is at zero.  It is clear
that the residuals are somewhat larger on average a few velocity
channels from the line centre, consistent with the example in
Figure~\ref{fig.linefit1}.

We therefore adopt a different strategy.  We create sets of pseudo
data by shifting the pattern of residuals about the peak of the best
fit curve while preserving their sequence and values.  For $n_{\rm
  boot} = 20$, say, we shift the residual pattern by one channel at a
time from $-10$ channels to $+10$ (skipping 0, of course) in a
deterministic manner at each iteration and refit.  Were the residual
pattern simply uncorrelated noise, this strategy would return similar
uncertainties in the fitted parameters as for random draws from the
pool of residuals, but preserving the intrinsic substructure of the
true emission line in this way yields a larger and more realistic
uncertainty in the fitted velocity.

\begin{figure}
\includegraphics[width=\hsize,angle=0]{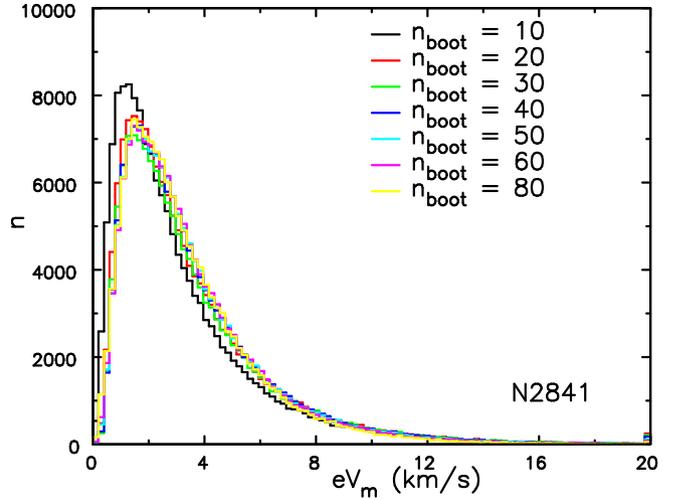}
\caption{The distributions of $eV_m$ from various numbers of
  bootstraps of fits to the data of NGC~2841 that included only the
  $h_3$ term. The differences for $n_{\rm boot}>10$ are minor.}
\label{fig.eVm}
\end{figure}

Obviously, the value of $n_{\rm boot}$ cannot be larger than the
number of channels and should usually be less.  We have experimented
with different numbers for NGC~2841, and show in Figure~\ref{fig.eVm}
that distribution of uncertainties changes very little for $n_{\rm
  boot} \ga 20$.  We have found similar distributions for two other
galaxies.  However, the median value of $eV_m \simeq 2.8\;$km~s$^{-1}$
in NGC~2841 is small compared with the likely turbulent velocity
spread in the gas of any star-forming galaxy.

\subsection{Bimodal line profiles}
\label{sec.bimod}
The Gauss-Hermite function (eq.~\ref{eq.profile}) can have two distinct
maxima near the line centre when $h_3$ and/or $h_4$ are
non-zero.\footnote{Formally, the function may also have minima and
  maxima in the wings of the profile, which are uninteresting because
  the $e^{-w^2/2}$ factor is already very small there.}  Two or more
peaks in the fitted line profile can arise because of noise or it
could imply streams of gas along the line of sight having distinct
velocities, which would violate one of the key assumptions implicit
when fitting a flow pattern to the data.

Program \mkm\ therefore flags fitted line profiles that are
significantly bimodal.  It determines whether the smooth function
${\cal L}(v)$ (eq.~\ref{eq.profile}) has multiple maxima and, if so,
whether any of the fitted secondary maxima are significant enough for
the line to qualify as bimodal.  It finds all the roots of
$\partial{\cal L}(v) / \partial v = 0$ and checks the sign of the
second derivative at the position of each root to locate the maxima in
the best fit line profile.  It discards any secondary maxima that are
less than 20\% of the greatest, the weaker of two maxima when two are
closely spaced, \ie\ when $\delta v < \sigma/2$, and cases where the
2nd maximum is scarcely more than inflexion in ${\cal L}(v)$.  To
identify this last case, the minimum between the two maxima must be
deeper than either 80\% of the lesser peak or 50\% of the average of
the two adjacent maxima.

When the number of surviving maxima is greater than one, a line is
flagged as bimodal by changing the sign of the uncertainty $eV_m$ if
$n_{\rm boot}>0$, or by changing the sign of the S/N otherwise.  Both
these quantities are intrinsically positive, so the flag is
unambiguous.  For example, the fitted profile in the middle panel of
Figure~\ref{fig.linefit1} is not bimodal, but the negative value of
$eV_m$ for the fit in the top panel flags the bimodality.

\citet{Oh08, Oh19} also try to identify bimodal line profiles, but
fit a second Gaussian to a side peak, describing the main peak as the
``bulk velocity''.  This is similar to our definition of the ``best
fit velocity'' $V_m$, which is that of the absolute maximum of the
smooth line profile (eq.~\ref{eq.profile}).  However, these authors
adopt the peak they judge to be the bulk velocity as the fitted
velocity, whereas we simply record the greatest peak and generally
discount lines that are flagged as bimodal.

\subsection{Omit $h_4$}
We caution that when fitting the six-parameter function
(eq.~\ref{eq.profile}) to low S/N data, the extra freedom in the
fitted function relative to a simple Gaussian could be used to fit a
spurious noise peak in addition to the true peak.
Figure~\ref{fig.linefit1} provides a good illustration of the dangers
of giving the fitting program more and more freedom to fit the data.
In this case, it is clear that the line fitting program is confused by
the spike near 820 km/s which could be real or a $3\sigma$ fluctuation
in the noise, and it reports a much smaller uncertainty $eV_m$, which
is negative to flag the line as bimodal for this fit.

This is one example to illustrate the undesirable consequences of
giving the fitting function too much freedom, and we recommend no more
than a five-parameter fit, with $h_4$ forced to be zero.

\section{Application to THINGS}
\label{sec.THINGS}
The data available on the THINGS website include moment 0, moment 1
and moment 2 maps, that are blanked outside the mask that the THINGS
team defined.  The unmasked ``standard'' cube, in the terminology of
\citet{Wa08}, is also available, but their posted moment maps are not
created from the standard cube.  We work with the ``natural weights''
data, which have slightly higher sensitivity and lower angular
resolution than do the ``robust weights'' data.

The standard cubes contain the intensities in the 21 cm line after an
almost complete data reduction, that includes cleaning of all sources
down to a level of 2.5 times the noise.  Two steps remain to be
performed in order to correct to the actual line fluxes: a rescaling
to make an approximate correction for the uncleaned sources caused by
the difference between the solid angle of the dirty and clean beams
\citep{JvM}, and a correction for primary beam attentuation.
\citet{Wa08} created the moment maps from the rescaled data cube, in
order that the flux in the moment 0 map be the best possible estimate.

However, correcting for both these effects will alter the noise, as
well as the signal, in every value in the cube, and a non-uniform
level of noise would complicate the fitting of line profiles.  The
THINGS team provide their ``standard'' cube, rather than the
``rescaled'' cube, at least in part because the noise is constant
throughout the cube \citep{Wa08}.  In fact, since primary beam
attenuation across a narrow frequency range is simply position
dependent, correcting for it has no effect on the shape of the line
profile in any pixel, and therefore does not change the fitted
velocity, as \citet{Wa08} point out.  However, rescaling to correct
for the dirty beam of uncleaned sources still present could, in
principle, mildly affect the shapes of the line profiles, although the
extent to which it could affect the fitted velocity is unclear.  With
this caveat, both dB08 and ourselves fit line profiles to the standard
cube.  The estimated noise in every pixel and velocity channel is that
given in Table 3 of \citet{Wa08}, and their values are in very good
agreement with our bi-weight esimates of the noise from the outer
layers of the cube (see \S\ref{sec.makemap}.1).

Many of the data cubes used in our analysis were also re-analyzed by
\citet{Po16}.  They smoothed the THINGS cubes to a common spatial
resolution of all their galaxies, rederived velocity maps from
Gaussian fits, discarding pixels having strongly skew line profiles,
and extracted rotation curves also using \rtc, without estimating
uncertainties.

\subsection{Masking}
\label{sec.setmask}
It is, of course, possible to attempt to fit a line to every pixel in
the cube, and to retain only those fits that are judged to have fitted
a real emission line.  We tried this at first, defining specific
criteria to discard a fit such as low signal/noise, fitted velocity
outside the range spanned by the channels, \etc \ However, some pixels
in the region far from the galaxy generally were accepted by our
adopted criteria, unless we made them so stringent that genuine
emission was also discarded.  The resulting map included a sparse
``haze'' of pixels having fitted ``lines'' that were perhaps emission
from gas clouds in the halo, or noise, which threw off the rotation
curve fitted to the outer parts of the map.  Thus it is best to devise
some other reason to discard isolated pixels.

This experience made it clear why it is common practice
\citep[\eg][]{Ro90, TA91, Wa08} to blank out parts of the data cube
before attempting to fit the real emission \citep[see][for a helpful
  explanation]{Da11}.  Many authors blank all values in the cube that
do not stand out from the noise, including velocity channels away from
the line emission.  However, we blank entire velocity columns at
pixels judged to have no emission from the target galaxy at any
velocity, but retain the full velocity spectrum in all remaining
pixels.  The full spectrum is needed for two reasons: fitting a line
profile requires a well-defined baseline, as was recognized by dB08,
and bootstrapping involves rearranging the residuals after fitting the
line profile.  Our mask is simply a binary toggle for each $(x,y)$
pixel, therefore.

We provide a very brief account of our procedure, which lacks the
sophistication of other packages.  For example, the SoFiA package
\citep{Serra15} seeks out coherent volumes within the 3D data cube
that are judged to contain real emission -- \ie\ the criteria include
coherence in velocity space, which we neglect.

\begin{figure}
\includegraphics[width=.8\hsize,angle=0]{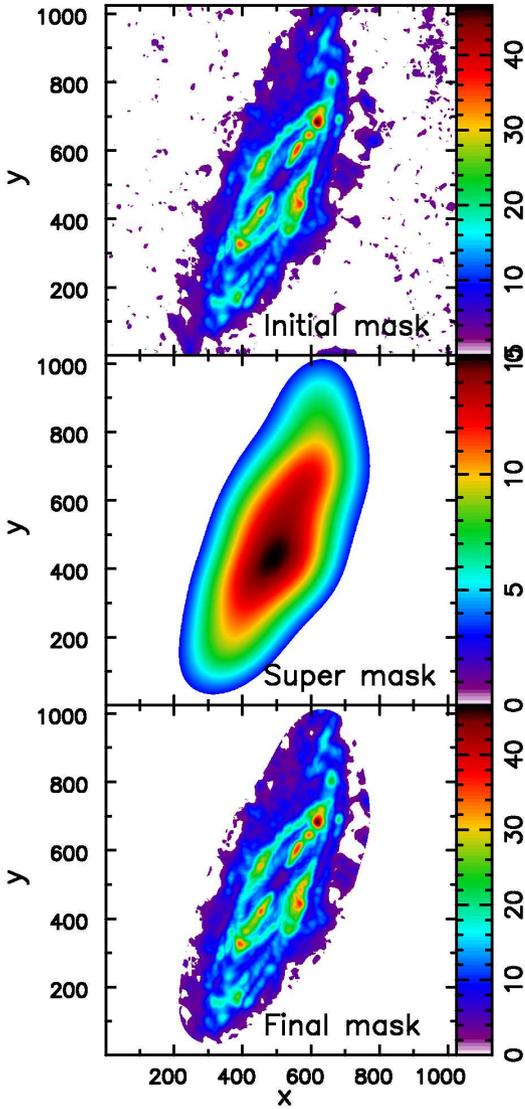}
\caption{Illustration of our procedure to make the mask using the
  natural weights THINGS data cube for NGC 2841.  The colored pixels
  in the top panel indicate the peak intensity, in units of ${\cal
    N}_s$, in the data cube smoothed by $30\arcsec$, while the white
  pixels are those below the threshold described in the text.  The
  middle panel is the result of heavily smoothing the image in the top
  panel and clipping away values below $3{\cal N}_s$, which we
  describe as the super mask.  The final map in bottom panel is the
  result of masking the top panel with the non-zero values in the
  super mask.  The final mask we employ selects only those pixels that
  have non-zero values in this last map.}
\label{fig.set_mask}
\end{figure}

Following \citet{Wa08}, we begin by smoothing the emission in each
channel map by convolving with a Gaussian of FWHM 30\arcsec.  (Since
the beam size in THINGS observations is already typically 10\arcsec,
the resulting resolution of the smoothed data is $\sim 32$\arcsec.)
We then calculate the global level of noise in the smoothed map,
${\cal N}_s$, again as the bi-weight estimate of a Gaussian spread of
the intensity in the outermost two planes of all six faces of the
smoothed data cube.  Then working over all $(x,y)$ pixels in turn, we
consider the pixel to have real emission if the smoothed intensity in
three consecutive velocity channels is greater than $2{\cal N}_s$, as
recommended by \citet{Wa08}.  The color scale in the top panel of
Figure~\ref{fig.set_mask} shows the peak intensity, in units of ${\cal
  N}_s$, in every pixel of the smoothed cube that is accepted by this
criterion.  The area containing the emission from the galaxy stands
out, but we need to do more to discard the many pixels away from the
galaxy that are also above the threshold.  The final masks created by
\citet{Wa08} also resulted from additional steps, that were not
described in their paper (de Blok 2020, private communication).

We therefore smooth the map in the top panel using a much broader
kernel of FWHM $\sim 71$ pixels, or $106\arcsec$ for this image,
obtaining the ``super mask'' map shown in the middle panel of
Figure~\ref{fig.set_mask}.  Here we have clipped away all pixels for
which this heavily smoothed intensity is $<3{\cal N}_s$ in all
velocity channels.  We could use the non-blanked channels of this
super mask as our mask, but we prefer to mask out any pixels within
this area that were previously masked in the top map, but have been
resurrected by smoothing.  This procedure leads to the final map shown
in the bottom panel.  Our mask becomes a binary acceptance of any
pixel that is colored in this map.  It excludes all pixels that are
well outside the galaxy, while generously including the entire area
containing emission from the galaxy.

Note that we discard the smoothed data cube after the mask is
determined, and follow standard practice to fit emission lines to the
data in the original unsmoothed cube.

\subsection{Extracting velocity and uncertainty maps}
\label{sec.getmaps}
We have remade velocity maps by the method described in
\S\ref{sec.makemap} from the standard data cubes of 18 of the 19
galaxies in the subsample of THINGS galaxies selected for rotation
curve analysis by dB08.  The one exception is M81 (NGC~3031), which
had been observed in two separate VLA pointings and for which the
available data cube is not in the standard form.

For each fitted galaxy, the Appendix presents the resulting velocity
maps and maps of the velocity uncertainties for each case.  Our
standard set of parameters for \mkm\ is: $ I_{n, \rm min} = 0.5$, $A_{\rm
  min} = 2$, $h_3$ included but $h_4$ excluded, $w_{\rm max} =
50\delta v$, $h_{\rm max} = 5$, and $n_{\rm boot}=20$.

\subsection{Extracting rotation curves}
\label{sec.DiskFit}
The \DFt\ package differs fundamentally from \rtc\ in that it tries to
fit a single idealized global model for the galaxy to the entire
observed velocity map.  In its simplest form, the model is a flat,
inclined disc in which the gas flows on circular orbits at speeds that
are tabulated at finite set of radii.  Such a model has five global
parameters, the systemic velocity of the galaxy, $V_{\rm sys}$, the
position of the centre of rotation $(x_{\rm cen}, y_{\rm cen})$, the
inclination, $i$, of the disc to the line of sight, and the position
angle, PA, of the major axis of the projected disc to due north.
\DFt\ can also allow that the flow in part of the disc is
intrinsically non-circular in a bar-like or oval distortion in the
same plane as the overall disc, or that the disc is intrinsically
warped in a simple parameterized manner.

\DFt\ proceeds by adjusting the parameters of the global fit to find
those that minimimize
\begin{equation}
\chi^2 = \sum_{\rm pixels} \left[ { V_{\rm line} - V_{\rm model} \over \sigma} \right]^2,
\label{eq.chisq}
\end{equation}
where $V_{\rm line}$ is the estimated velocity from the Doppler shift
of the line at each pixel and $\sigma$ is the adopted uncertainty in
$V_{\rm line}$.  The predicted velocity at each pixel, $V_{\rm
  model}$, is computed by interpolation between a set of ellipses at
which the circular speed, and any non-circular streaming motions, are
tabulated.  For each set of global parameters, the optimal choices for
these velocities is first determined by matrix inversion
\citep[see][for details]{SS07}.

Our line fitting procedure (\mkm, \S\ref{sec.makemap}) returns an
estimate of the uncertainty, $\sigma_{\rm line}$, in the fitted
velocity at each pixel that, if used as the denominator in
eq.~(\ref{eq.chisq}), would have the desirable effect of giving high
relative weight to pixels with small velocity uncertainties.  Since
the median uncertainty in the fitted velocity is typically $<
3\;$km~s$^{-1}$, and some are a small fraction of this, the pixels
having the smallest values of $\sigma_{\rm line}$ would dominate the
global fit to an excessive extent, however.  We think it unlikely that
the line-of-sight velocity at a particular location can be determined
with a precision that is a small fraction of the level of turbulence
in the ISM within a galaxy.  We therefore choose a non-zero value for
$\Delta_{\rm ISM}$, which is a global constant that we add in
quadrature to the values of $\sigma_{\rm line}$ at each pixel and set
$\sigma = (\sigma_{\rm line}^2 + \Delta_{\rm ISM}^2)^{1/2}$ in
eq.~(\ref{eq.chisq}).  The larger the value of $\Delta_{\rm ISM}$ the
less strongly the pixels are weighted towards those where the velocity
is well determined.

It is important to note that not only have we propagated the
uncertainty in the fitted velocity in each pixel into the global
minimization, but \DFt\ returns uncertainties in all the fitted global
parameters and estimated orbit speeds at each radius that are
determined by a modified bootstrap method proposed by \citet{SZS10}.
Their method assumes that we see the galaxy as a disc in projection
and attempts to preserve coherent line-of-sight velocity residuals
arising from spiral-arm streaming.  The algorithm deprojects the
residual pattern to face on, rotates it through a random angle, and
rescales it in radius before reprojecting, and then adds the resulting
new residual pattern to the best fit model to create a pseudo data set
for a new fit.  For fits that include a bar flow, or oval distortion,
we adopt the same procedure separately for the regions within the bar
and outside it.  This approach allows the user to estimate
uncertainties in all the global parameters and the circular speed at
each radius that not only take account of statistical errors from the
observations, but also allows for spiral arm streaming, turbulence and
other possible departures of the galaxy from the simple model adopted.
Furthermore, we examine the spreads in the values of global parameters
for possible covariances that may indicate inadequacies in our model.

These capabilities contrast strongly with the tilted-ring approach
used in \rtc, which is widely employed and was adopted by dB08, in
particular.  To wit: \DFt\ assumes a generally flat disc, with a fixed
$V_{\rm sys}$, centre, PA, and $i$ for the whole galaxy, whereas all
these quantities can optionally be allowed to vary from ring to ring
in \rtc.  Furthermore, uncertainties in the circular speed generally
quoted by users of \rtc\ are estimated separately for each ring and
take no account of uncertainties in the systemic velocity, projection
geometry, or global inadequacies of the tilted-ring model.

In the present application to the THINGS data, we discard pixels for
which the fitted line is flagged as bimodal, and estimate
uncertainties from 200 bootstrap iterations.  We have adopted values
in the range $1 \leq \Delta_{\rm ISM} \leq 8\;$km~s$^{-1}$, choosing
smaller values generally for dwarf galaxies.  Small changes to
$\Delta_{\rm ISM}$ have little effect on the fit, but we found it was
sometimes necessary to employ larger values when $\chi^2$ became
excessive.  Although we do not attach any meaning to this statistic,
and therefore do not quote it, values of $\chi^2 \ga 5$ per degree of
freedom were a warning that the fit was being driven by the pixels
having small $\sigma$ in equation (\ref{eq.chisq}).  Raising the value
of $\Delta_{\rm ISM}$ in these cases naturally reduced $\chi^2$ and
improved the behaviour of \DFt.

As noted above, \DFt\ tabulates the rotation curve at a set of radii,
or more precisely ellipse semi-major axes, and the model predictions
are interpolated between the tabulated ellipses for each pixel in
equation~(\ref{eq.chisq}).  Thus the sampling of the rotation curve
can be much coarser, typically 1 - 4 beam widths, in contrast to every
half-beam width that is preferred in \rtc.  The beam sizes in the
individual THINGS galaxies are given in Table 3 of \citet{Wa08}, and
the FWHM are in the range 5\arcsec to 15\arcsec.

\begin{table*}
\begin{tabular}{@{}lccccccc}
Galaxy & $\Delta_{\rm ISM}$ & $V_{\rm sys}$ & $i$ &  PA & centre RA & centre $\delta$ & bar PA in \\
name & km s$^{-1}$ & km s$^{-1}$ & deg & deg & pixels & pixels & disc plane: deg \\
NGC 925 & 5 & $548.7 \pm 3.0$ & $61.5 \pm 2.8$ & $-75.7 \pm 1.0$ 
& $518.5 \pm 5.2$ & $516.9 \pm  4.2$ \\
NGC 2403 & 8 & $132.8 \pm 1.1$ & $61.8 \pm 1.7$ & $124.1 \pm 0.5$
& $1020.8 \pm 4.0$ & $1008.9 \pm 2.9$ \\
NGC 2841 & 5 & $631.0 \pm 1.9$ & $73.7 \pm 1.3$ & $148.6 \pm 1.5$
& $508.1 \pm 1.0$ & $513.2 \pm 1.5$ \\
NGC 2903 & 8 & $552.9 \pm 1.1$ & $64.4 \pm 0.9$ & $203.1 \pm 0.8$ 
& $510.7 \pm 0.6$ & $513.8 \pm 1.4$ & $230.6 \pm 11.4$ \\
NGC 2976 & 5 & $3.2 \pm 0.8$ & $58.4 \pm 0.8$ & $-36.5 \pm 0.3$ 
& $514.5 \pm 1.2$ & $515.8 \pm 1.1$ & $155.6 \pm 4.9$ \\
IC 2574 & 5 & $45.4 \pm 4.3$ & $47.7 \pm 9.9$ & $57.2 \pm 1.7$ 
& $544.2 \pm 26.6$ & $524.1 \pm 17.4$ \\
NGC 3198 & 5 & $683.9 \pm 0.7$ & $71.1 \pm 0.8$ & $215.6 \pm 0.5$ 
& $511.4 \pm 0.8$ & $512.9 \pm 1.0$  \\
NGC 3521 & 8 & $789.4 \pm 4.2$ & $70.9 \pm 1.2$ & $-20.0 \pm 0.7$ 
& $506.2 \pm 1.8$ & $513.8 \pm 2.8$  \\
NGC 3621 & 8 & $727.4 \pm 1.2$ & $64.7 \pm 0.9$ & $-15.9 \pm 0.4$
& $507.8 \pm 1.5$ & $511.2 \pm 1.8$ \\
NGC 3627 & 8 & $710.7 \pm 2.8$ & $56.7 \pm 2.4$ & $172.9 \pm 2.1$ 
& $508.9 \pm 1.7$ & $513.8 \pm 1.9$ & $60.2 \pm 14.9$ \\
NGC 4736 & 8 & $310.1 \pm 0.5$ & $31.7 \pm 5.1$ & $-55.1 \pm 1.1$
& $512.6 \pm 0.5$ & $513.4 \pm 0.4$ & $241.9 \pm 7.7$ \\
DDO 154 & 1 & $375.1 \pm 0.4$ & $64.1 \pm 1.3$ & $225.6 \pm 0.5$ 
& $503.2 \pm 1.5$ & $510.5 \pm 1.7$  \\
NGC 4826 & 8 & $411.1 \pm 1.7$ & $64.3 \pm 2.0$ & $122.1 \pm 2.0$ 
& $514.5 \pm 0.8$ & $517.1 \pm 0.9$  \\
NGC 5055 & 5 & $502.0 \pm 1.4$ & $65.8 \pm 0.6$ & $100.2 \pm 0.6$
& $516.1 \pm 1.0$ & $513.2 \pm 0.4$ \\
NGC 6946 & 6 & $51.0 \pm 2.1$ & $36.4 \pm 3.8$ & $243.0 \pm 1.0$
& $513.6 \pm 5.3$ & $517.9 \pm 5.5$ \\
NGC 7331 & 8 & $816.4 \pm 4.0$ & $76.0 \pm 0.8$ & $169.9 \pm 0.5$ 
& $511.8 \pm 0.6$ & $513.7 \pm 1.9$  \\
NGC 7793 & 4 & $228.4 \pm 0.3$ & $43.9 \pm 4.3$ & $-65.0 \pm 0.9$ 
& $506.1 \pm 0.7$ & $516.6 \pm 1.2$ & $50.8 \pm 5.3$  \\
\end{tabular}
\caption{The 17 galaxies in our sample and parameters of the best fit
  models in each case obtained by applying \DFt\ to our re-derived
  velocity map.  The values of $V_{\rm sys}$, $i$, and PA may be
  compared with average values given in Table~2 of \citet{deB08}, who
  employed \rtc\ to fit tilted ring models to their own velocity maps
  derived from the same data cubes.}
\label{tab.summary}
\end{table*}

\section{Results}
\label{sec.results}
Of the 18 THINGS galaxies for which we made new velocity maps, we
fitted rotation curves using \DFt\ to 17; fits to the remaining
galaxy, NGC~2366, were not believable (see \S\ref{sec.NGC2366}).  The
rotation curves from \DFt\ are also presented in the Appendix,
together with maps of residuals after subtracting the best fit model,
and notes on the individual galaxies.  Table~\ref{tab.summary} gives
the global parameters of the fitted models, and their uncertainties,
as well as the adopted value of $\Delta_{\rm ISM}$ in each case.

Our fitted rotation curves are generally, with 2 or 3 exceptions, in
good agreement with those already published by dB08 who derived their
velocity maps independently and employed \rtc\ to estimate the
circular speed at each radius.  However, our estimates of the
uncertainty in the circular speed generally differ from theirs.  The
rotation curves reported in the later study by \citet{Po16} that
included many of the same galaxies were also generally similar, but
these authors did not present uncertainties in the orbit speeds at
each radius.  \citet{SZS10} had previously used an earlier version
\DFt\ to extract rotation curves for five THINGS galaxies.  Since we
here rederive velocity maps using a different algorithm, we include
these five galaxies in the present study.

We have employed a flat disc model in every case except NGC~2841, for
which we fitted a mild, parametric warp, and NGC~5055.  It is well
known that warps are common in the outer discs of many galaxies; see
\citet{Sell13} for a review of both the observational evidence and
theory.  The first of the three summary rules for warps deduced from
observational data by \citet{Brig90} is that the warp generally starts
at $R_{25}$, the radius at which disc surface brightness in the blue
band falls below 25 mag per square arcsec.  This finding is in
agreement with theory \citep[\eg][]{SS06}, which also predicts that
the disc within $\sim 4$ disc scale lengths is expected to be rigid
enough to stay flat.

However, an axisymmetric flat disc model was not a good fit to the
part of the flow well within $R_{25}$ for five galaxies: NGC~2903,
NGC~2976, NGC~3627, NGC~4376, and NGC~7793, which we therefore fitted
with a bar or oval distortion.  In these same cases, \rtc\ finds that
the PA and/or inclination of the ``disc'' varies strongly in the
region of the bar or oval flow, which is a natural consequence of
insisting that an intrinsically elliptical flow pattern be modelled as
motion in a circle.  We are confident that such apparently
pathological twists in the inner disc are an artifact of \rtc, and
argue that the inner disc does in fact have a fixed projection
geometry and instead gas is streaming in an elliptical flow pattern in
a non-axisymmetric potential over part of the radial range.

Our fits to most galaxies do in fact extend well beyond $R_{25}$, yet
we continue to find acceptable fits with a flat disc model, with
little evidence in the velocity residuals for a warp, with the
exceptions of NGC~2841 and NGC~5055 that we discuss below.  The
absence of a large warp in the outer disc is consistent with the
findings of dB08 for which their fitted values of PA and/or $i$
generally vary by $\la 10^\circ$ from those in the inner disc.  This
statement remains true for the cases with bars when the region where
we fit a bar is discounted, but does not hold for NGC~2841 and
NGC~5055.  Aside from these last two cases, we find that forcing a
flat disc model does not lead to systematic discrepancies in the
fitted circular speeds returned by \DFt\ and \rtc.

Our fits to the velocity map of NGC~2841 were improved by allowing a
mild parameterized warp model, as discussed in \S\ref{sec.N2841}.  The
paramterization of the warp adopted in \DFt\ was unsatisfactory for
NGC~5055, however, and instead we found that an acceptable fit could
be made by allowing separate fixed-projection geometries of the inner
and outer parts of the disc, with the boundary at a semi-major axis of
375\arcsec, which is very close to the value $R_{25}=
378$\arcsec\ given in the RC3 \citep{dVCB}.  In fact, this can be
achieved in a single fit without dividing the data by allowing for
non-circular flows in the inner disc with the projected geometry set
by the outer disc, as we report in the Appendix.  Fitting the two
parts of the flow with such a ``bar model'' is nothing more than a
device to capture two flat, intrinsically circular flow patterns that
are misaligned in the different parts of the galaxy.  The values for
the global parameters of NGC~5055 given in Table~\ref{tab.summary} are
for the fit to the inner disc only, while the fitted values for the
outer disc are $i = 59.3\pm1.1$ and PA$ = 92.7\pm0.6$, with the centre
held fixed at the same position as fitted for the inner disc.

\subsection{Uncertainties in the fitted parameters}
Uncertainites in the global parameters, as well as the circular speed
at each radius, were estimated from 200 bootstrap iterations in
\DFt\ using in all cases the procedure recommended by \citet{SZS10},
as described above (\S\ref{sec.DiskFit}).

We have not computed the covariance matrix, and instead present
figures in the Appendix showing distributions of values of the fitted
paramaters for each bootstrap iteration, with the best fit parameters
marked in red.  Our uncertainties in the parameters quoted in table
\ref{tab.summary} are computed from the spreads of these values in the
coordinate directions.  These figures are generally reassuring because
the best fit model usually lies near the centres of these
distributions and there are few pronounced correlations between the
parameters.  The most obvious exception is that the systemic velocity
of some galaxies correlates with the fitted position of the centre
along the major axis, especially for dwarf galaxies for which the
rotation curve rises slowly.  This is to be expected, since it is
well-known that the velocity map of a hypothetical galaxy having an
exactly linearly rising rotation curve cannot constrain either the
position of the centre along the major axis, and consequently the
systemic velocity, or the galaxy inclination \citep{Bo78}.  All the
THINGS galaxies rotate differentially to some degree, which breaks
these exact degeneracies, but a vestige of them lingers in the
parameter correlations from the bootstrap iterations, and is
particularly pronounced for IC~2574 (Figure~\ref{fig.IC2574}).

It should be noted that one of the reasons that we rejected all our
attempts to fit a model to the data for NGC~2366 was that the ``best
fit'' models were outliers in all the parameter covariance plots.
Rearranging the large velocity residuals in this case seemed always to
lead to fits for which the global parameters differed from the best
fit values in a lop-sided sense.  We interpreted this as a red flag to
indicate that the best fit should not be accepted, as we note below in
the discussion of this galaxy in \S\ref{sec.NGC2366}.  The ability to
provide this kind of warning diagnostic to flag doubtful fits is
unique to our analysis methods.

\begin{figure}
\includegraphics[width=\hsize,angle=0]{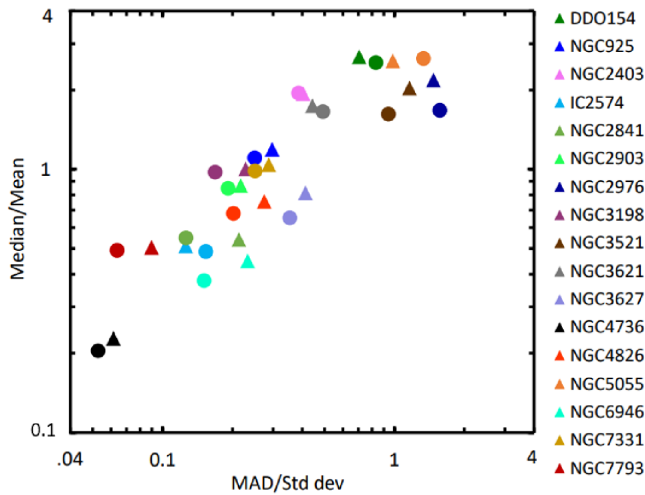}
\caption{Comparison of velocity uncertainties from \rtc\ and \DFt\ as
  described in the text.  For each galaxy, we compute the ratio of the
  uncertainties from \rtc\ to those from \DFt, and characterize the
  distribution of values in four ways.  The circles mark the median as
  a function of median absolute deviation (MAD), while the triangles
  give the mean as a function of standard deviation.  The data from
  the figures in \citet{deB08} were kindly provided in digital form by
  Erwin de Blok.}
\label{fig.uncerts}
\end{figure}

\subsection{Circular speed uncertainties}
The figures in the Appendix present estimated circular speeds, as well
as their uncertainties, as functions of radius both from our work
(points with error bars) and those from dB08 (red lines with grey
shading).  Figure~\ref{fig.uncerts} summarizes how the two sets of
uncertainty estimates compare.  To make this Figure, we computed the
ratio of the error estimates reported by dB08 divided by those given
by \DFt.  Since dB08 sample the rotation curve more densely, we
interpolate their error estimates to the radius of the \DFt\ point,
omitting values where the two RCs do not overlap and where we included
a bar/oval.  We then derive four measures of the distribution of
ratios for each galaxy: the mean, standard deviation, median, and
median absolute deviation (MAD), as shown in the Figure.

In some galaxies (\eg\ IC~2574, NGC~4736, NGC~6946 and NGC~7793) our
uncertainties are substantially larger than those reported by dB08,
which is usually attributable to a significant uncertainty in the
inclination of the disc to the line-of-sight, a source of uncertainty
that \rtc\ ignores.  In other cases (\eg\ DDO~154 and NGC~5055), our
uncertainties are substantially smaller than theirs.  In the remaining
cases there is approximate agreement, although in NGC~2841 and in
NGC~3521 we disagree over just part of the radial range.  Our
substantially smaller uncertainties in the outer part of the rotation
curve for NGC~3521 probably stem from our elimination of pixels for
which the line profile was flagged as bimodal, as the data cube
appears to include emission from a, possibly infalling, gas stream
having a line-of-sight velocity below that of the circular speed in
the disc mid-plane.

\subsection{Three examples}
Here we highlight our findings for three galaxies that illustrate
the power of our methods.

\subsubsection{DDO 154}
\label{sec.DDO154}
This dwarf galaxy is extremely well fitted by our flat, axisymmetric
disc model, since the velocity residuals, Figure~\ref{fig.DDO154}, are
generally small.  The covariance plots in the bottom panel of the same
figure reveal that each global parameter of the best fit model, red
symbols, is near the centre of the distribution of values from the 200
bootstrap iterations, and the projection geometry is tightly
constrained.  Strong degeneracies between the location of the centre,
the inclination, and systemic velocity can arise when fitting velocity
maps of dwarf galaxies, but in this case they are mild, with just weak
correlations in the expected sense between $V_{\rm sys}$ and the
$(x,y)$ position of the centre.  dB08 allow $i$ and PA to vary with
radius, but remark that they ignore the variation near the centre and
in the outer parts because both regions suffer from sparse data, and
extrapolate from the almost constant and better constrained values at
intermediate radii. Their table 2 gives $V_{\rm sys} =
375.9\;$km~s$^{-1}$, and average values of $i = 66.0^\circ$ and PA$=
229.7^\circ$, in good agreement with our values in
Table~\ref{tab.summary}.  Our rotation curve agrees well with that of
dB08, but our estimated uncertainties in the circular speed are
substantially smaller than theirs.

In summary, our work strengthens the evidence from previous studies
that this galaxy has a particularly well-determined rotation curve
with very little wiggle room for comparisons with predictions.

\subsubsection{NGC 2841}
\label{sec.N2841}
As fits to the velocity map of this galaxy were unsatisfactory when
\DFt\ assumed a flat disc, we allowed for a gradual mild warp.  The
parameterized warp fitted by \DFt\ allows for a change of both
inclination and position angle with radius that varies in a quadratic
fashion from zero change at the inner radius $R_{\rm warp}$ to a
maximum at the edge.  Thus there are three additional global
parameters to be fitted: the radius at which the warp starts, $R_{\rm
  warp}$, and the maximum changes in ellipticity $w_{\epsilon, \rm
  max}$ and PA $w_{\phi, \rm max}$.

The results are displayed in Figure~\ref{fig.N2841}.  The covariance
plots are generally well behaved, but many values of $R_{\rm warp}$
are clustered at the low end of the range, because we created a wall
in the $\chi^2$ function to prevent $R_{\rm warp}$ from straying below
the radius of the second fitting ellipse.  The other parameters of the
mild warp are: $w_{\epsilon, \rm max} = 0.02 \pm 0.04$, which
corresponds to an increase in inclination of $\sim 3^\circ$ from the
centre to the edge, while the PA of the major axis increases from the
centre to the edge by $w_{\phi, \rm max} = 13.6^\circ \pm 2.4^\circ$.

Table 2 of dB08 gives $V_{\rm sys} = 633.7\;$km~s$^{-1}$, and average
values of $i = 73.7^\circ$ and PA$= 152.6^\circ$.  The angles in our
Table~\ref{tab.summary} apply to the inner disc.  The change in PA
that we fit is about the same as that reported by dB08, but those
authors report an increase in $i$ of some 8$^\circ$, which is larger
than we find.

Our rotation curve is in excellent agreement with the fit reported by
dB08 and our uncertainties are similar to their estimates in the outer
disc and a little larger than theirs in the inner disc.

\begin{figure}
\includegraphics[width=.8\hsize,angle=0]{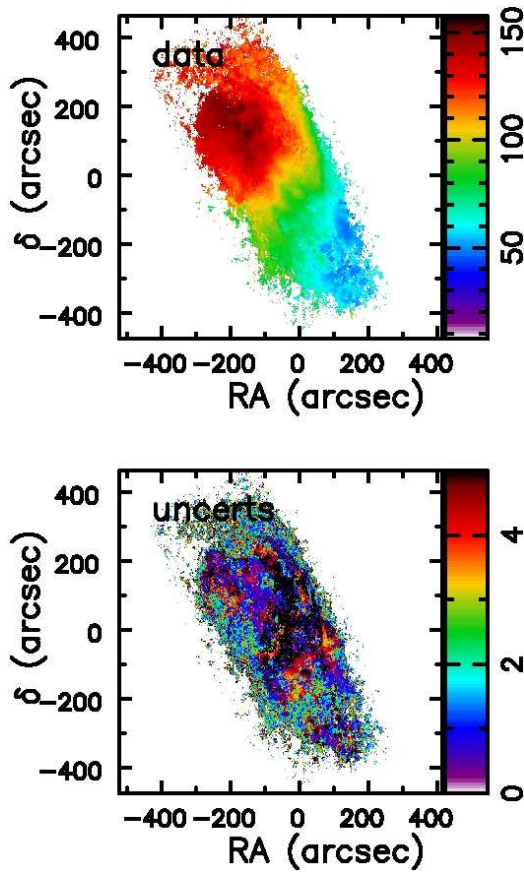}
\caption{Our velocity map (above) with uncertainties (below) for NGC
  2366.}
\label{fig.N2366}
\end{figure}

\subsubsection{NGC 2366}
\label{sec.NGC2366}
We were unable to find a credible model that fitted the velocity map
of the dwarf galaxy NGC~2366.  The upper panel of
Figure~\ref{fig.N2366} manifests a clear velocity gradient roughly
aligned with the major axis of the HI emission, but the iso-velocity
contours do not appear to have the characteristic pattern of flow in
flat, axisymmetric disc.  Note also in the lower panel that many
pixels near the centre of our velocity map have uncertainties
$>5\;$km~s$^{-1}$.

The rotation curve reported by dB08 rises to nearly 60~km~s$^{-1}$
around 250\arcsec, but then drops below 40~km~s$^{-1}$ by
420\arcsec, which is a steeper than Keplerian fall off.  Similar
shapes were also reported in previous studies \citep{Sw99, Hu01}.
\citet{Oh08}, on the other hand, obtained a quite different fit to the
same data cube, which is also reported in dB08.  They suggested the
circular speed continues to rise for $R>300\arcsec$, finding an
essentially flat, axisymmetric disc in the outer parts.  They obtained
this result by applying \rtc\ to a map of the ``bulk velocity'' in the
line profiles as described in their paper.

Our attempts to use \DFt\ to fit a flat, axisymmetric model to our
full map yielded a bizarre rotation curve, which was scarcely improved
by allowing an oval distortion.  A strongly warped disc model yielded
a better fit with a more reasonable rotation curve, but with large and
coherent residuals.  However, the rotation centre was far from the
centre of the map, and the deprojected data filled only slightly more
than a semi-circle.  This circumstance precluded the use of the
bootstrap method proposed by \citet{SZS10}, which assumes a round disc
seen in projection in which spiral streaming motions are the main
failing of the model.  We were therefore forced to use the earlier
bootstrap method proposed by \citet{SS07} that attempted to preserve
the patchwise coherence of the velocity residuals.  This method left
the ``best fit'' model as an outlier from the cluster of values in the
covariance plots in most parameters, which is another red flag, and we
therefore doubted this fit also.

Furthermore, discarding some of the data in the outer parts of the map
caused substantial changes to the fitted position of the centre, the
overall inclination, the magnitude of the warp, and the estimated
rotation curve.  We were unable to find a significant subset of the
data that would yield consistent values for any of these parameters.

We therefore conclude that the gas in this dwarf galaxy does not
follow a simple flow pattern in even a twisted disc and do not present
a fitted model for this galaxy.

\section{Discussion and conclusions}
\label{sec.discuss}
Any meaningful comparison between a model prediction and observations
requires that the data with which the model is compared have
well-defined uncertainties so that the likelihood that the model
matches the data can be assessed.  In the case of rotation curves
derived from high resolution 21cm line observations of nearby
galaxies, the uncertainties in the fitted circular speed have not, as
yet, been derived in a statistically meaningful manner, and therefore
the success of galaxy formation models in predicting the observed data
cannot be quantified.

As a further step towards the goal of a meaningful comparison, we have
presented a new standalone tool, \mkm, to extract 2D velocity maps
from 3D data cubes of spectral line data.  Unlike previous approaches
to extract velocity maps from high resolution spectral data cubes, our
method estimates the uncertainty in the fitted velocity in each pixel
that takes into account the clumpy and turbulent nature of the ISM in
the observed galaxy.  The estimated velocity uncertainties can then be
propagated into the procedure to fit a model to the velocity map.  The
modern methods we reviewed in the introduction that attempt to fit a
model directly to the 3D data cube, without the intermediate step of
making a velocity map, are inefficient when applied to high
spatial resolution data, and also have not, thus far, attempted to
estimate uncertainties in a well-founded manner.

From previous work \citep{SS07, SZS10}, we have made public a model
fitting program, \DFt, that is an improvement over the usual
tilted-ring fitting procedure to extract rotation curves because it
estimates uncertainties in the fitted circular speed that take account
of large-scale turbulence, spiral arm streaming motions, as well as
global uncertainties in the position of the centre, the systemic
velocity, and the inclination and position angle of the galaxy to the
line of sight.  In the absence of uncertainties in the observed
velocities to be fitted by the model, this code simply assumed a
constant uncertainty in the fitted velocity in each pixel in the map.
But our new map making tool provides uncertainties in the fitted
velocity at each pixel that can now be propagated into the model
fitting.

Here we have applied both the new tool, \mkm, together with \DFt\ to
the publicly available data cubes of 17 galaxies from the THINGS
survey \citep{Wa08}.  Our re-analysis these data has generally
confirmed the rotation curves derived by \citet{deB08}, although we
find a few differences.  We have adopted flat disc models for most of
the galaxies, allowing a bar-like or oval distortion in the inner
parts of five galaxies, whereas the tilted ring analysis by dB08
generally fitted the same data with strong twists to the disc plane.
Our model for NGC~2841 included a mild warp, consistent with the
findings of dB08.  The velocity map of NGC~5055 is also consistent
with an abrupt warp that begins near $R_{25}$, and we have fitted this
galaxy with separate inner and outer flat discs having different
projection geometries.

More importantly, we believe our analysis provides better-founded
uncertainty estimates that reflect both the uncertainties in the data
and in the projection parameters as well as non-circular streaming
motions.  In this sense, we have taken a step towards the ultimate
goal of statistically valid uncertainties.  We do claim that our
quoted uncertainties in model parameters and rotation curves are more
reasonable than those in previous work, but do not claim to have
evaluated them with full statistical rigour.  Our uncertainties are
sometimes a few times greater than those given by dB08, and sometimes
just a fraction of their values.

We highlight two example galaxies from our study to illustrate which
nearby galaxies provide useful comparisons with model predictions, and
which do not.  We find that the data on DDO~154 are extremely well
fitted by a simple, flat axisymmetric flow pattern having tightly
constrained projection parameters and that dB08 in fact overestimated
the uncertainties in the fitted circular speed.  On the other hand, we
could not find a credible fit to the data from NGC~2366.  Therefore we
would argue that DDO~154 provides an excellent test case for
comparison with galaxy formation models, but that nothing useful could
be learned from including NGC~2366 in such a study.

dB08 fitted mass models to their rotation curves, weighting each data
point by the inverse square of their estimate of its error.  A similar
fitting procedure to our very similar derived rotation curves would,
to be sure, be differently affected by our different uncertainties,
perhaps leading to slightly different best fit models.  However, our
key point is that having more realistic uncertainties would enable us
to assess the likelihood that the data matches any fitted model,
something that no previous study that we are aware of has attempted to
quantify.  We therefore defer this major task to a separate paper.

\section*{Acknowledgements}
We thank Erwin de Blok for being very responsive to our questions and
for providing rotation curve data and other data in digital form. We
also thank an anonymous referee for a helpful and detailed report.
JAS gratefully acknowledges the continuing hospitality of Steward
Observatory. KS acknowledges support from the Natural Sciences and
Engineering Research Council (NSERC) of Canada.

\section*{Data availability}
The tools \mkm\ and \smk\ will be added to the \DFt\ package website:
{\tt
  https://www.physics.queensu.ca/Astro/people\-/Kristine\underline{\phantom{a}}Spekkens/diskfit/}.
The raw data cubes are avaliable on the THINGS website ({\tt
  https://www2.mpia-hd.mpg.de/THINGS/Data.html}).  Our maps, models,
rotation curves and uncertainties are available on request.

\def\skip#1{ \etal\ }
\def\PhD{PhD thesis.}
\def\pnas{{\it Proc.\ Nat.\ Acad.\ Sci.\ (USA)}}

\appendix
\section{Fits to individual galaxies}
\label{sec.notes}

Here we present new velocity maps, with uncertainties, created by
\mkm\ for 17 THINGS galaxies, together with model and residual maps
and rotation curves fitted to the new velocity maps using \DFt.  The
best fit global parameters are given in Table~\ref{tab.summary}, and
the uncertainties and error bars on the fitted circular speeds are
derived from 200 bootstrap iterations.  For each galaxy, we also
present plots to test for covariances between the fitted global
parameters and provide a short narrative discussion of each case.

\begin{figure*}
\includegraphics[width=.82\hsize,angle=0]{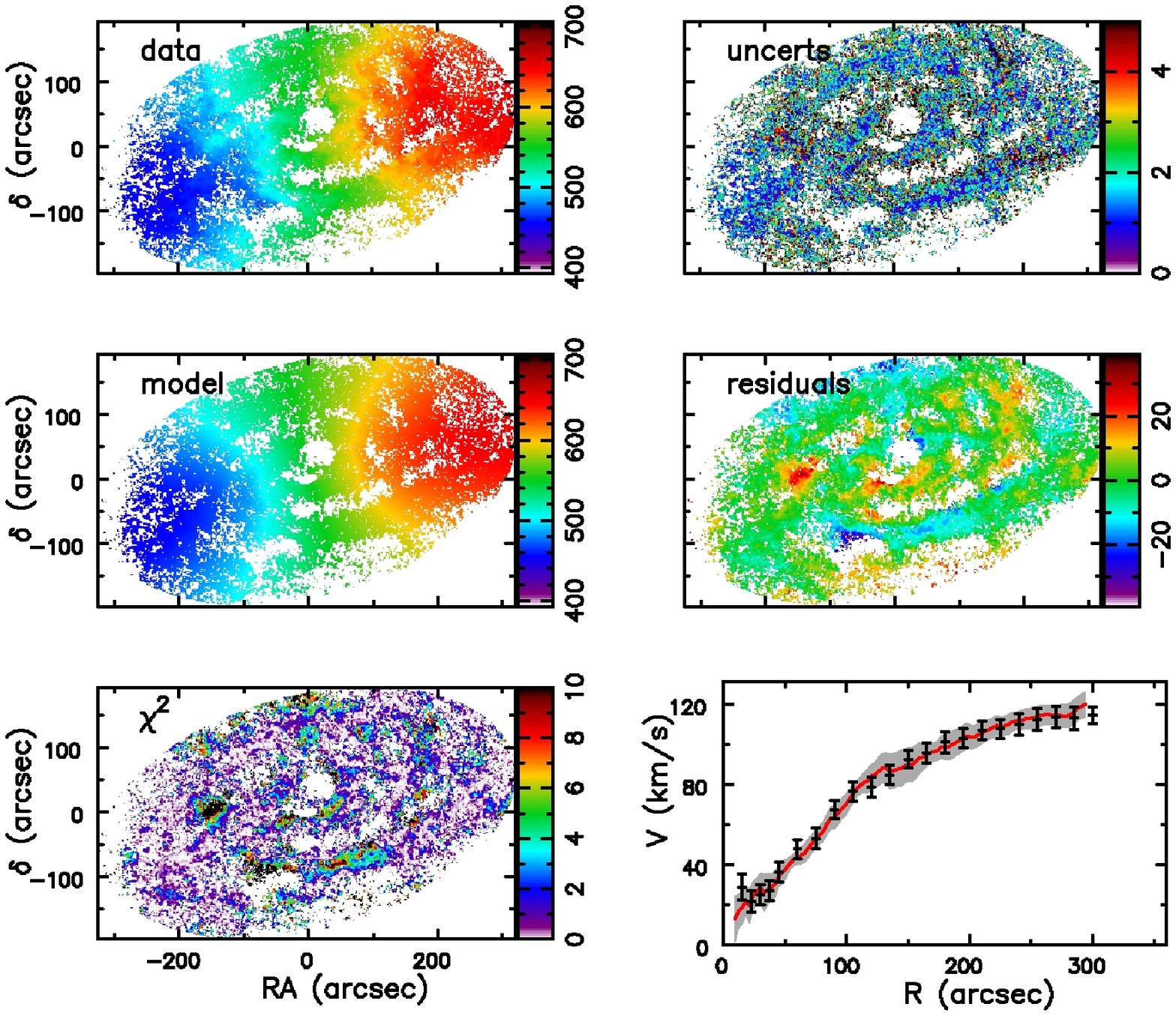}
\includegraphics[width=.62\hsize,angle=0]{cvN925.ps}
\caption{NGC 925 Top left: the velocity map extracted from the THINGS
  ``standard cube'', with line-of-sight velocity in km~s$^{-1}$ colour
  coded as indicated, and top right the uncertainty in the fitted
  velocity, also in km~s$^{-1}$.  2nd row left: the best-fit model
  from \DFt, and right the map of residual velocities after
  subtracting the fitted model from the data.  Third row left, map of
  the values of $[(V_{\rm line} - V_{\rm model}) / \sigma]^2$.  Third
  row right: the black points with error bars show the fitted rotation
  curve, with the error bars resulting from 200 bootstrap iterations.
  The bottom panel shows the covariance of the global parameters;
  $V_{\rm sys}$ strongly correlates with $x_{\rm cen}$ because the
  major axis is close to the $x$-axis. The best fit projection
  parameters with uncertainties are given in Table~\ref{tab.summary}.
  The red line in the rotation curve panel is that presented in dB08
  with their estimates of the uncertainties indicated by the shading.}
\label{fig.N925}
\end{figure*}

\subsection{NGC 925}
This dwarf galaxy has a simple velocity map that is reasonably well
fitted as an axisymmetric, inclined disc.  The black points with error
bars in the third row right panel of Figure~\ref{fig.N925} indicate
our fitted rotation curve and its uncertainties, which are in
reasonable agreement with the same quantities estimated by dB08,
marked by the red line and grey shading.  The kink in the red line
occurs at the radius where their \rtc\ analysis finds a discontinuity
in the disc inclination, which we do not reproduce in our flat disc
fit.  The residuals map has few large values but does not reveal any
tell-tale coherent features that would suggest a non-axisymmetric
potential.  We conclude that our axisymmetric model is a reasonable,
albeit imperfect, fit to the data.

\begin{figure*}
\includegraphics[width=.8\hsize,angle=0]{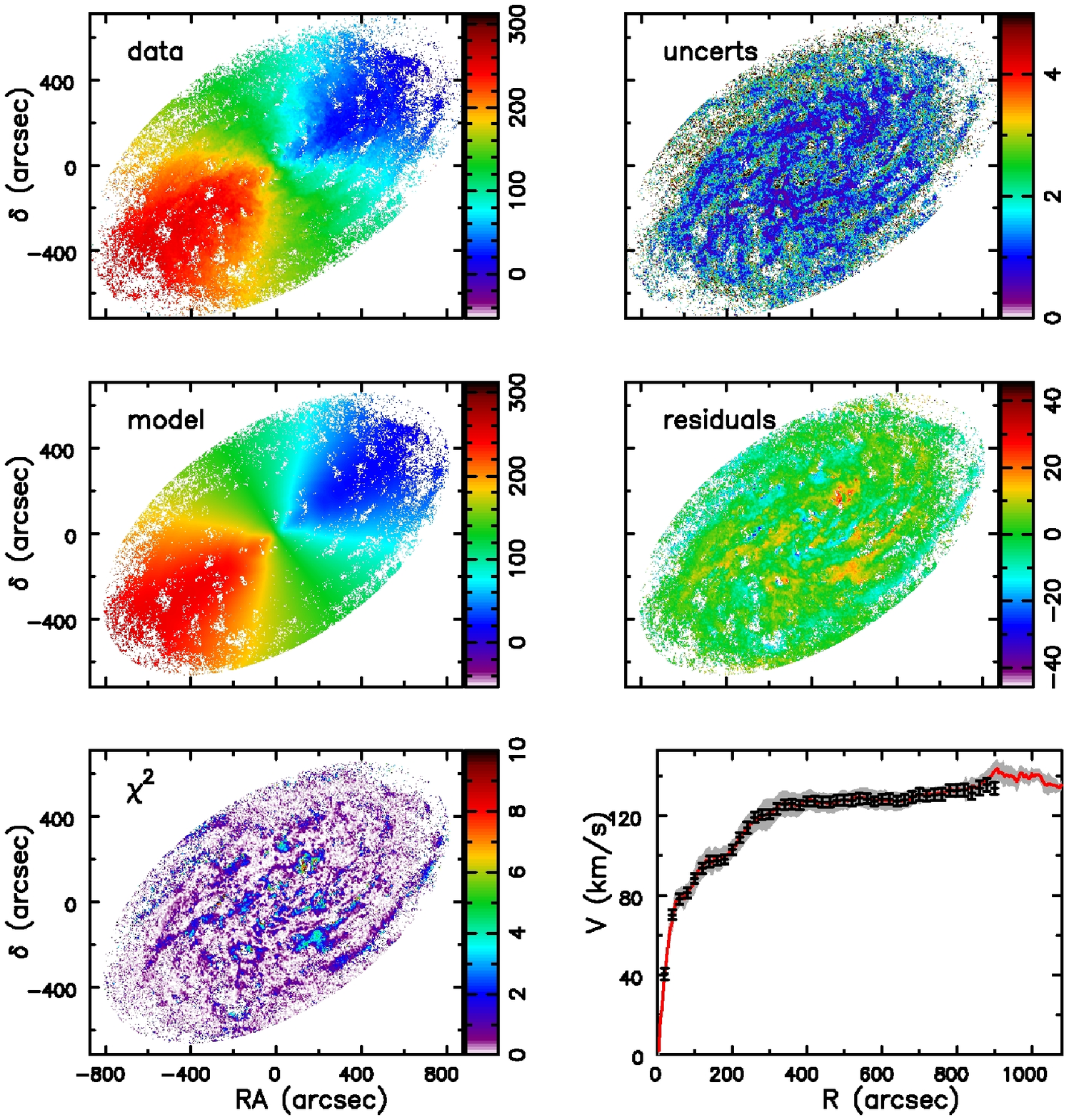}
\includegraphics[width=.68\hsize,angle=0]{cvN2403.ps}
\caption{As in Figure~\ref{fig.N925}, but for NGC 2403.}
\label{fig.N2403}
\end{figure*}

\subsection{NGC 2403}
Our flat, axisymmetric disc model fits the data from this galaxy very
well, as shown in Figure~\ref{fig.N2403}.  The rotation curve is in
excellent agreement with that from dB08, and also \citet{SZS10},
although our uncertainties tend to be smaller than those quoted by the
THINGS team.  \citet{SZS10} were unable to find evidence for a
significant non-axisymmetric component in the potential.  Our fit
excluded gas emission beyond 900\arcsec\ because at least some of it
caused the rotation curve to drop at those radii.  We suspect that
this behaviour was due to gas at anomalous velocities that was not
eliminated by our mask.

\begin{figure*}
\includegraphics[width=.9\hsize,angle=0]{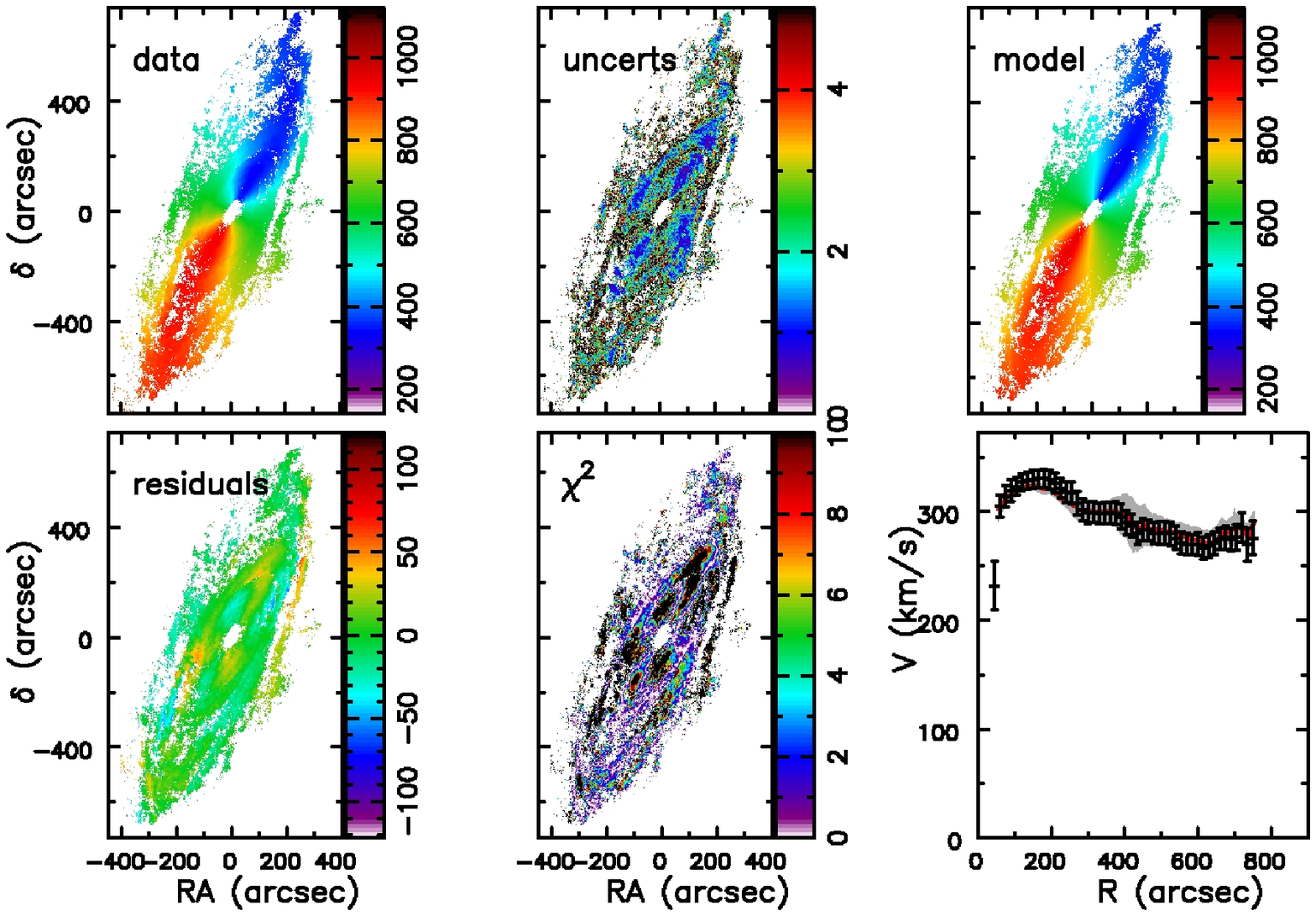}
\includegraphics[width=.9\hsize,angle=0]{cvN2841.ps}
\caption{As in Figure~\ref{fig.N925}, but for NGC 2841.}
\label{fig.N2841}
\end{figure*}

\subsection{NGC 2841}
As fits to the velocity map of this galaxy were unsatisfactory when
\DFt\ assumed a flat disc, we allowed for a gradual mild warp, as
described in \S\ref{sec.N2841}.

\begin{figure*}
\includegraphics[width=.9\hsize,angle=0]{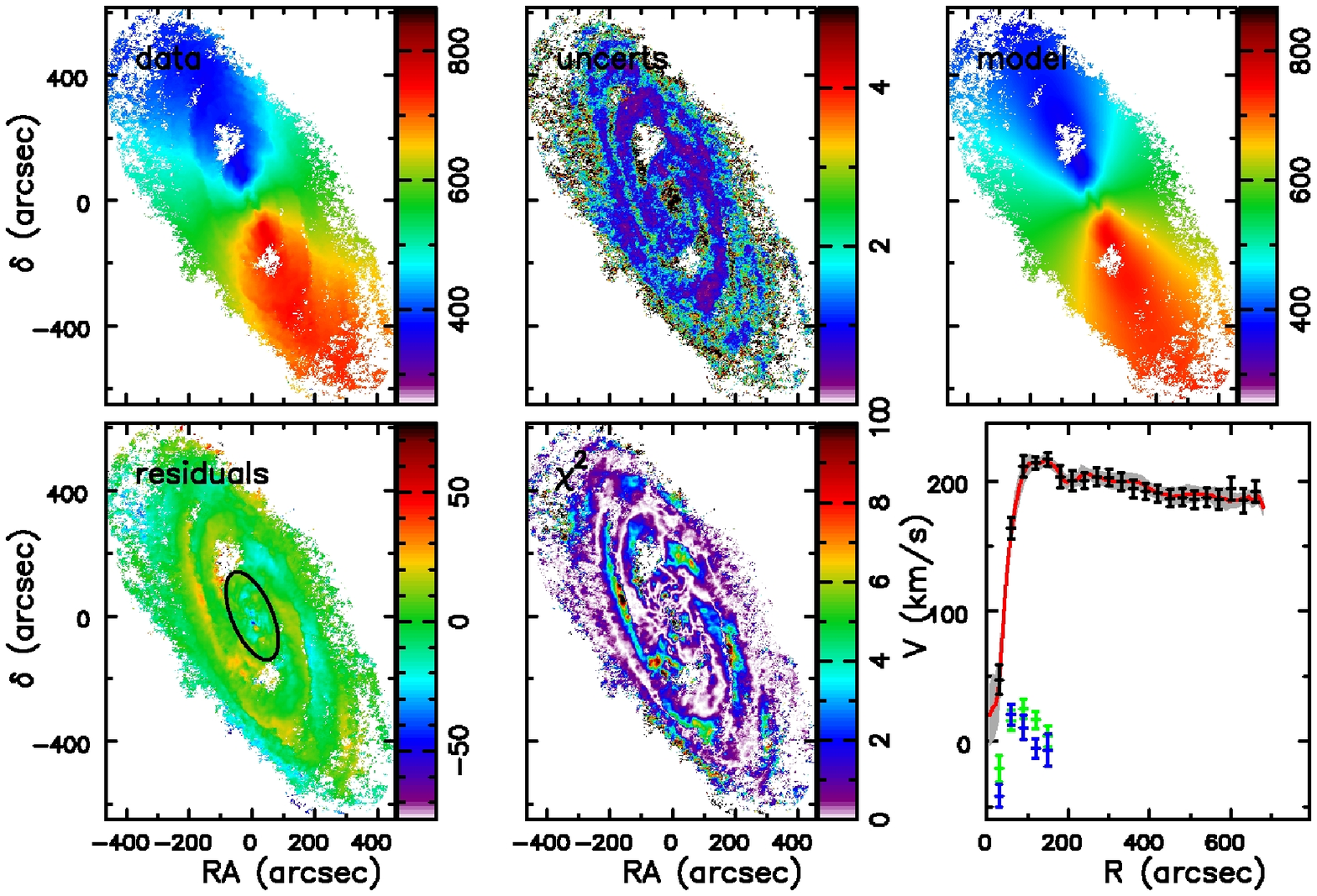}
\includegraphics[width=.8\hsize,angle=0]{cvN2903.ps}
\caption{As in Figure~\ref{fig.N925}, but for NGC 2903.  The green and
  blue points with error bars indicate the fitted radial and azimuthal
  velocity components of the bar flow, which is fitted inside the
  ellipse drawn on the residuals map.}
\label{fig.N2903}
\end{figure*}

\subsection{NGC 2903}
Since the $3.6\,\mu$m {\it Spitzer} image shows the disc of this
galaxy to be barred, we have included a bar flow to the fitted model
in the inner part of the disc.  The fitted radial and azimuthal bar
flow velocities \citep[][]{SS07} are indicated, respectively, by the
green and blue points with error bars.  The bar flow velocities are
small compared with the mean orbital speed, which agrees well with
that estimated by dB08 who found, not surprisingly, that axisymmetric
fits in the bar region required the PA and inclination to vary.

\citet{SZS10} also used an earlier version of \DFt\ and included a bar
flow but to 60\arcsec\ only.  We find similar results as they did when
we limit the non-asisymmetric part to the same region, but here allow
for a somewhat more extensive bar flow.  Even so, the residual
velocity map manifests a coherent anti-symmetric ring, which is at a
similar location as a ring in the HI column density and may possibly
have some small radial velocity.

\begin{figure*}
\includegraphics[width=.9\hsize,angle=0]{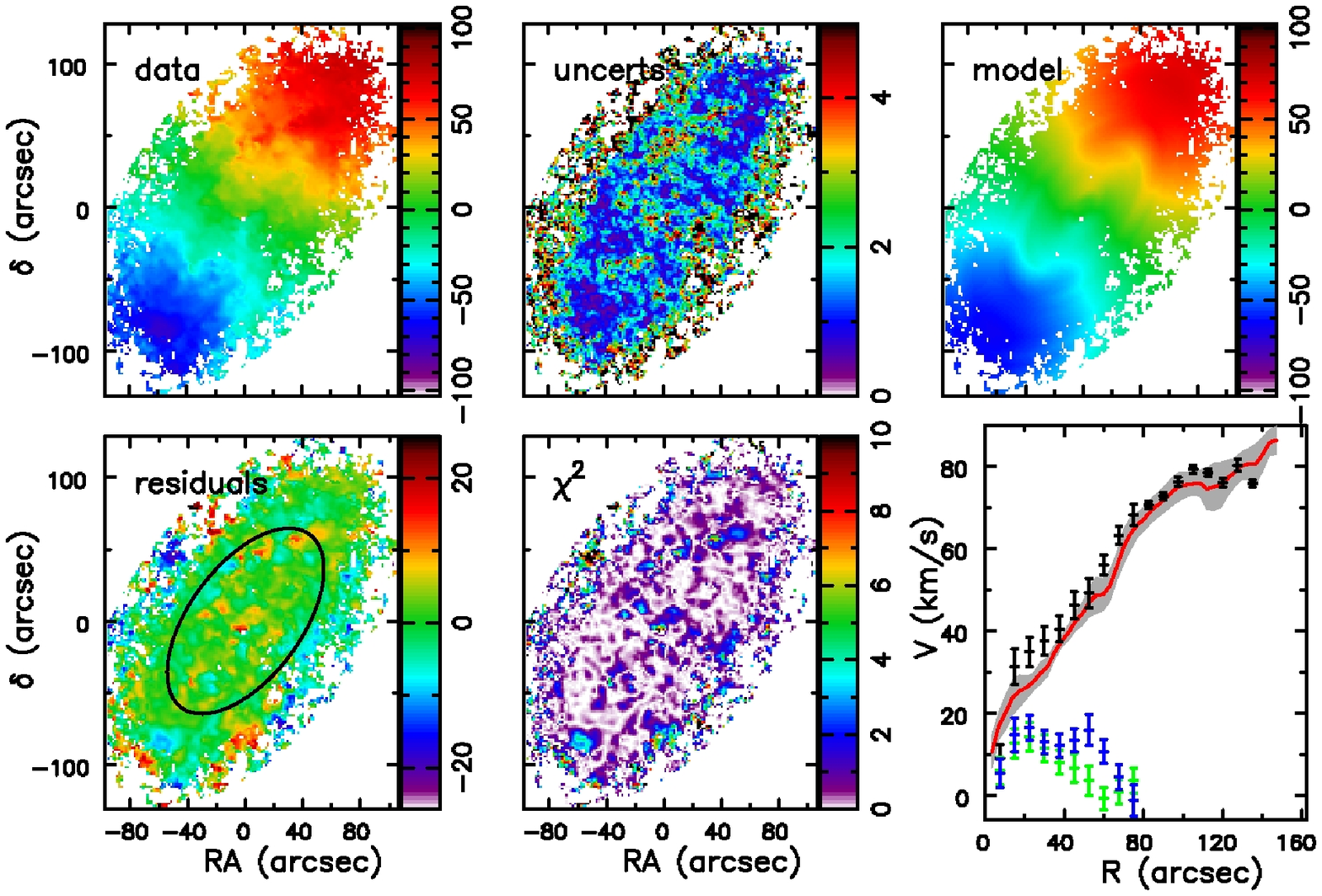}
\includegraphics[width=.8\hsize,angle=0]{cvN2976.ps}
\caption{As in Figure~\ref{fig.N2903}, but for NGC 2976.}
\label{fig.N2976}
\end{figure*}

\subsection{NGC 2976}
\citet{SS07} first showed that other velocity maps for this galaxy
were better fitted with a bar flow, and this remains true for the
THINGS data.  As in our earlier work, and in \citet{SZS10}, fitting
for an inner oval leads to higher estimated mean speeds over the
radial range of the oval than were estimated by dB08.  The THINGS team
again reported large apparent changes in the fitted projection
geometry that were caused by adopting an axisymmetric flow pattern
where the flow is better fitted by oval streamlines in a flat disc.

\begin{figure*}
\includegraphics[width=.75\hsize,angle=0]{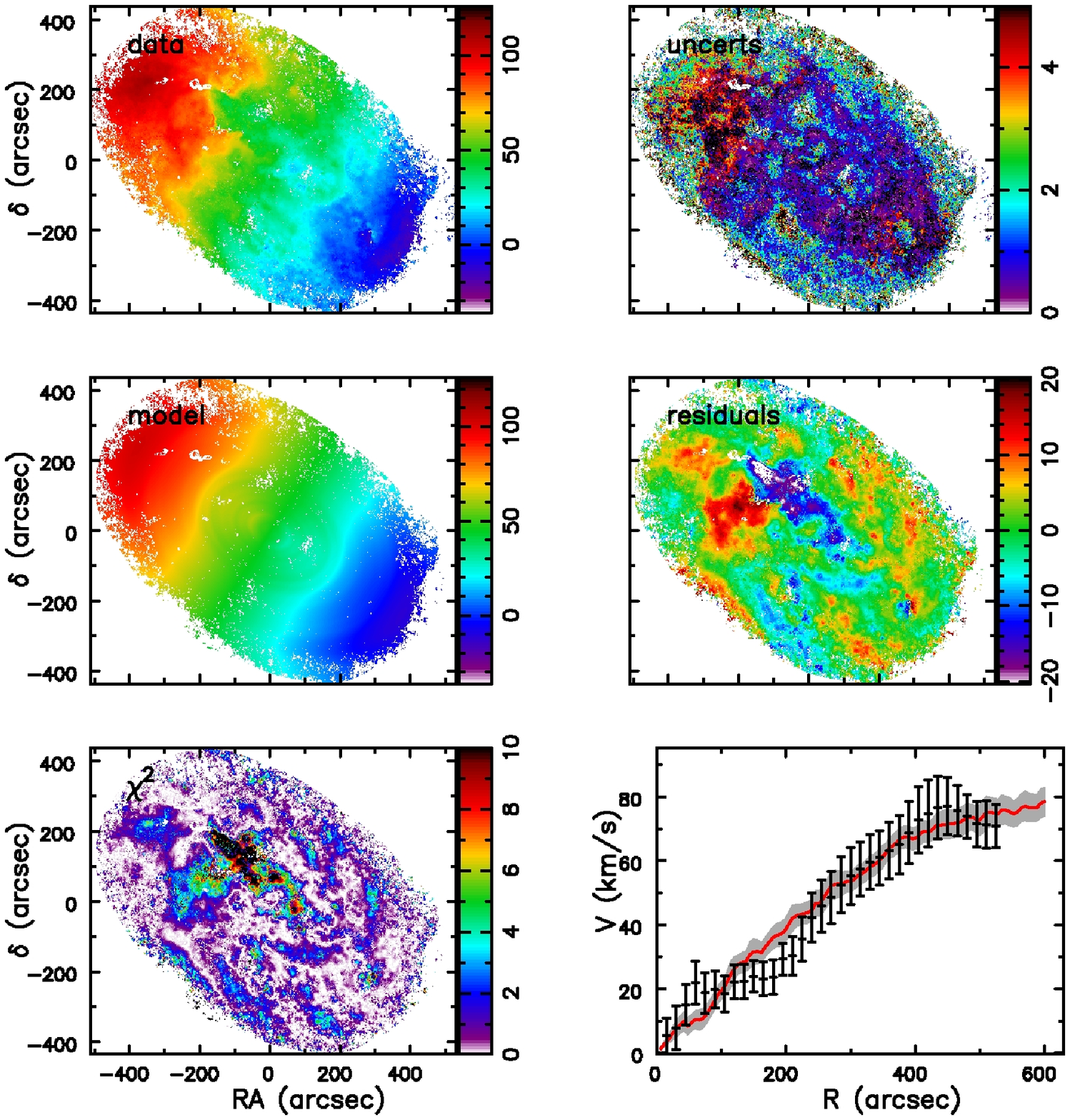}
\includegraphics[width=.68\hsize,angle=0]{cvIC2574.ps}
\caption{As in Figure~\ref{fig.N925}, but for IC 2574.}
\label{fig.IC2574}
\end{figure*}

\subsection{IC 2574}
This galaxy is a member of the M81 group and is surrounded by a number
of substantial HI clouds \citet{So20}.  It is not well fitted by our
axisymmetric model Figure~\ref{fig.IC2574}.  Some of the velocity
residuals are large, but there is no tell-tale pattern in them to
suggest a non-axisymmetry in the flow and allowing a bar/oval
component to the flow made only a marginal improvement.  Our best fit
inclination and systemic velocity (Table~\ref{tab.summary}) are in
good agreement with the values obtained by \citet{So20} from their
direct fits to their low resolution 3D data cube, but both differ
significantly from the values given in table 2 of dB08, which are
$i=53.3^\circ$ and $53.1\;$km~s$^{-1}$.  Also our rotation curve is in
rough agreement with that reported by dB08 and better agreement with
that derived from the lower resolution data by \citet{So20}.  Our
estimates of the uncertainties, which largely stem from an almost
10$^\circ$ uncertainty in the fitted inclination
(Table~\ref{tab.summary}), are much greater than those suggested by
the THINGS team.

\begin{figure*}
\includegraphics[width=.9\hsize,angle=0]{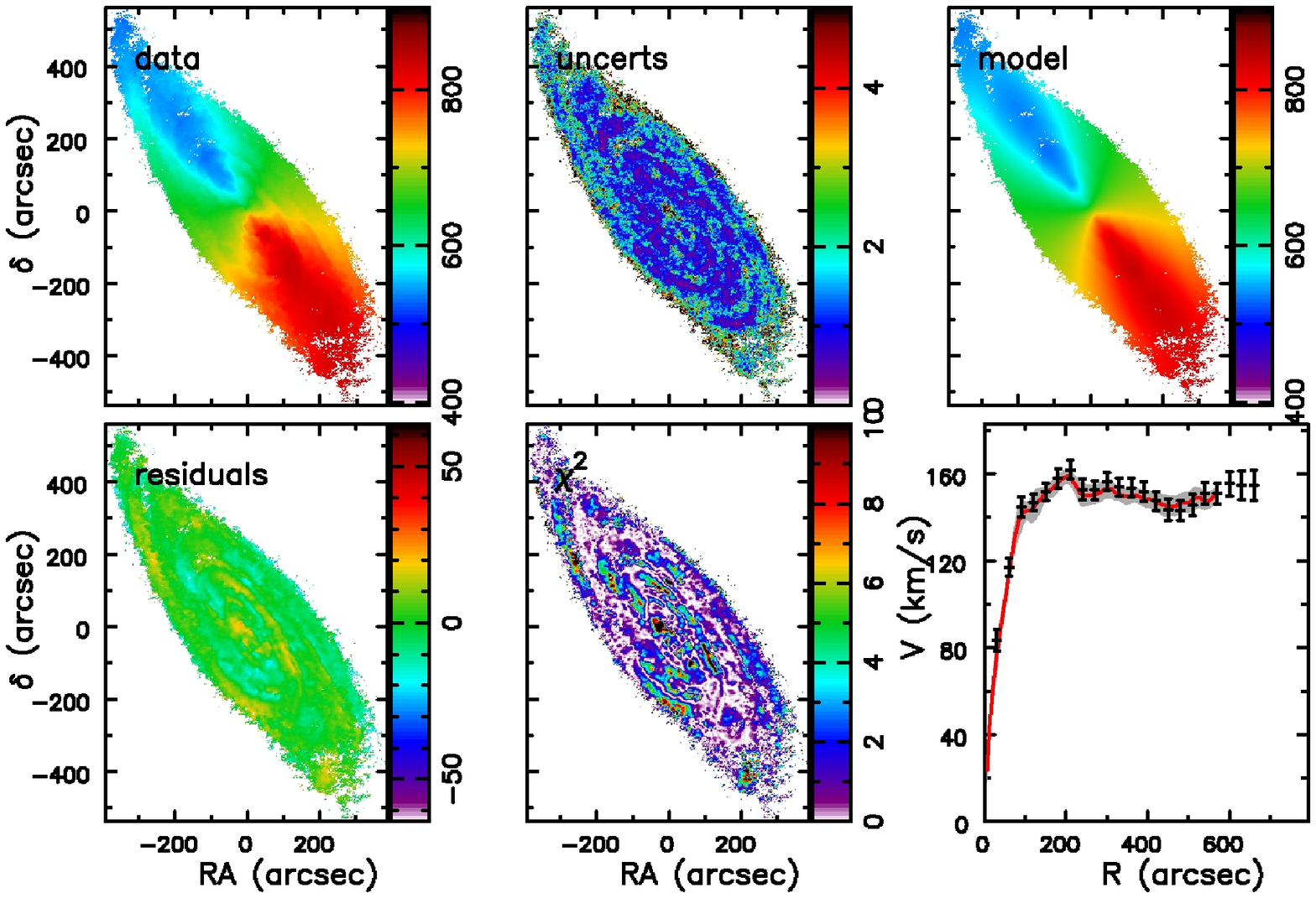}
\includegraphics[width=.68\hsize,angle=0]{cvN3198.ps}
\caption{As in Figure~\ref{fig.N925}, but for NGC 3198.}
\label{fig.N3198}
\end{figure*}

\subsection{NGC 3198}
A flat axisymmetric model fits the data from this galaxy very well, as
is evident from Figure~\ref{fig.N3198}.  Our fitted rotation curve and
uncertainties are in good agreement with those estimated by dB08.
\citet{SZS10}, who used an earlier version of \DFt\ to search for
non-axisymmetic distortions, failed to find any evidence for a
distortion in the mid-plane potential of this well-studied galaxy.

\begin{figure*}
\includegraphics[width=.9\hsize,angle=0]{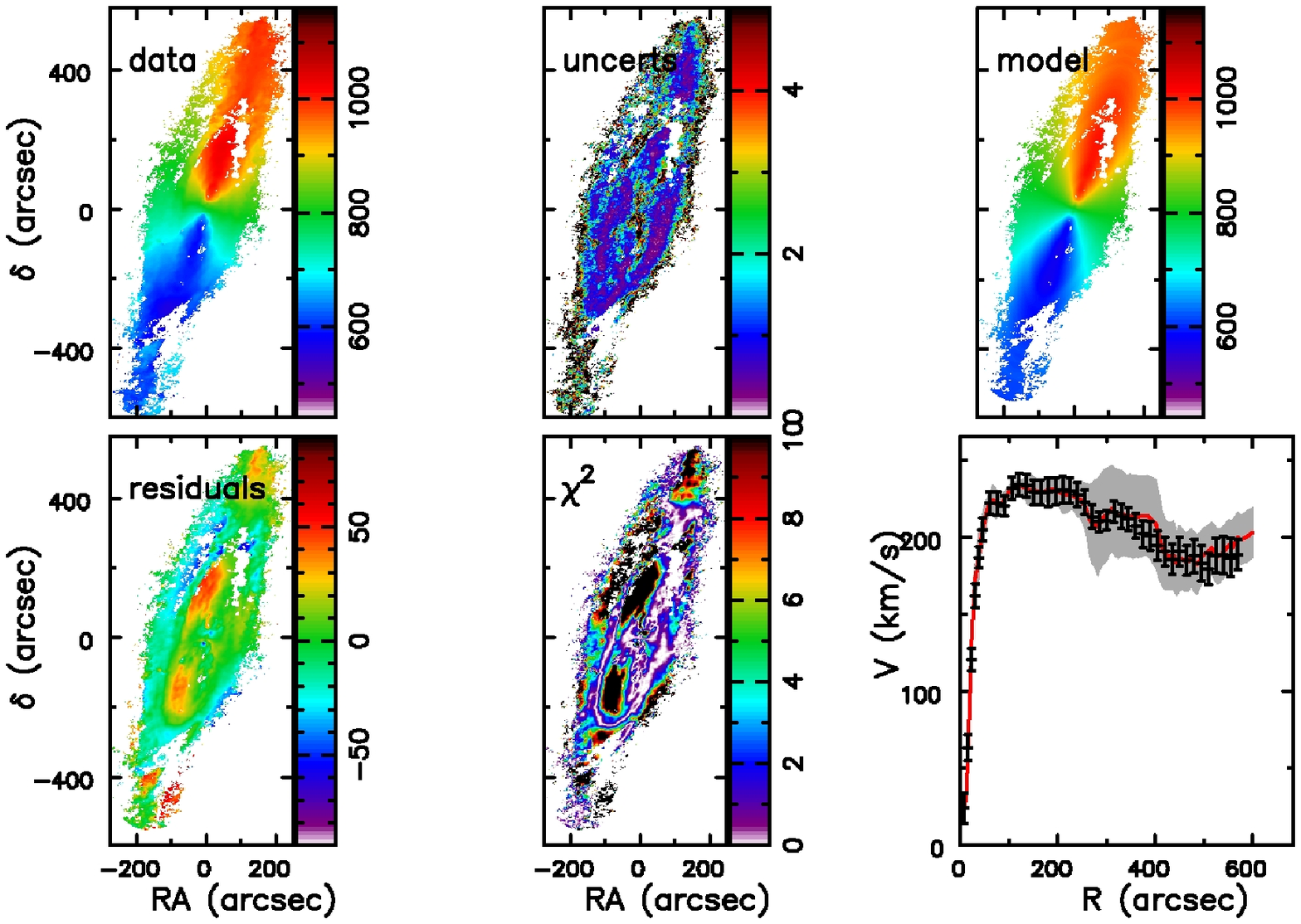}
\includegraphics[width=.68\hsize,angle=0]{cvN3521.ps}
\caption{As in Figure~\ref{fig.N925}, but for NGC 3521.}
\label{fig.N3521}
\end{figure*}

\subsection{NGC 3521}
Large parts of the velocity map of this galaxy have bimodal line
profiles, possibly suggesting that emission is both from the disc and
a stream of extra-planar gas.  We have excluded pixels having bimodal
line profiles and fit an axisymmetric model that has a declining
rotation curve.  Our fit, Figure~\ref{fig.N3521}, is in reasonable
agreement with that from dB08, but our uncertainties are smaller,
probably because we excluded pixels having bimodal profiles.  The
axisymmetric fit is hardly satisfactory because the velocity
difference maps reveal areas of large coherent residuals, although we
were unable to discern a coherent pattern to them.

\begin{figure*}
\includegraphics[width=.9\hsize,angle=0]{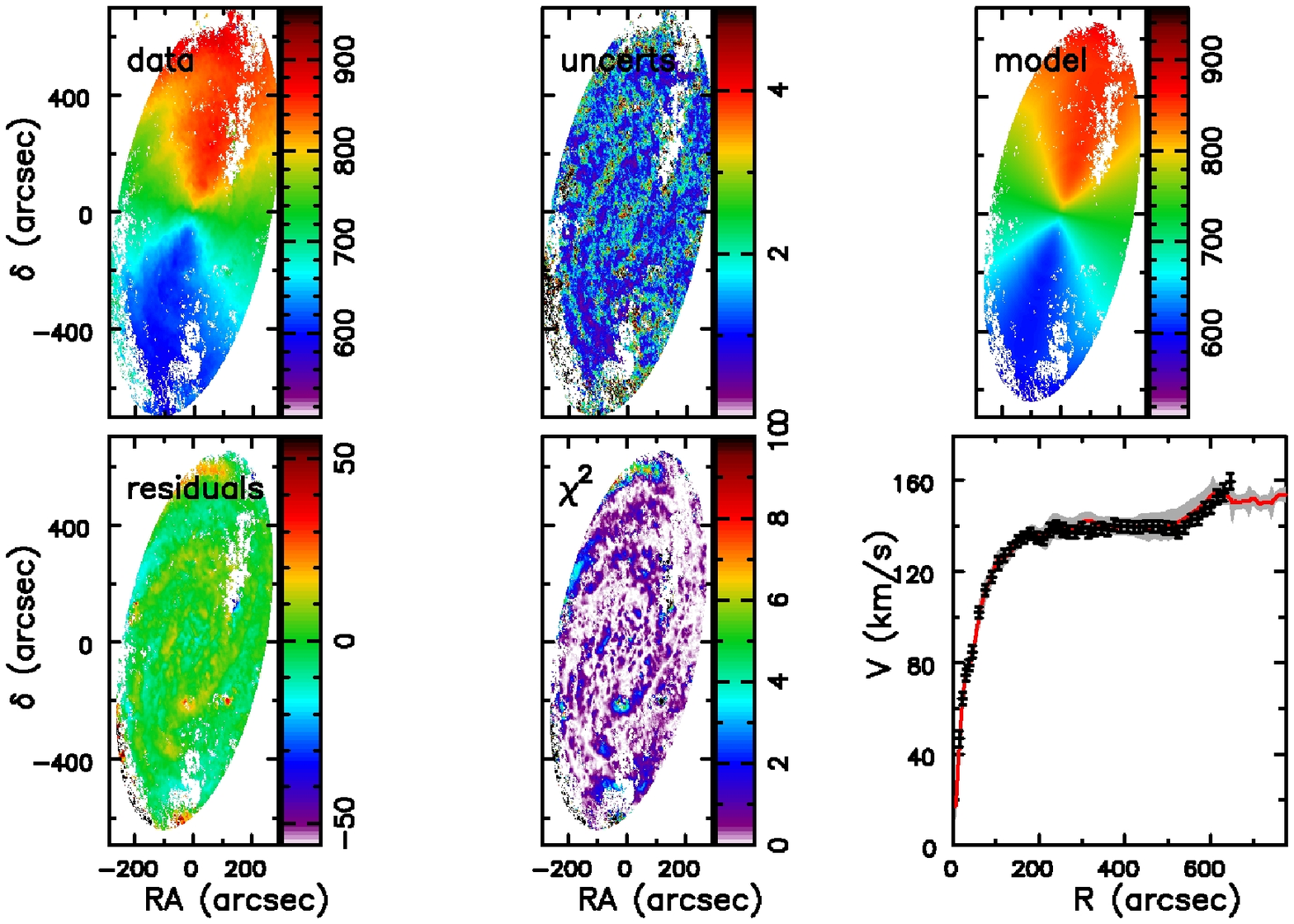}
\includegraphics[width=.68\hsize,angle=0]{cvN3621.ps}
\caption{As in Figure~\ref{fig.N925}, but for NGC 3621.}
\label{fig.N3621}
\end{figure*}

\subsection{NGC 3621}
This galaxy is well fitted by a simple axisymmetric disc flow model,
as shown in Figure~\ref{fig.N3621}.  The rotation curve agrees well
with that given by dB08, although our uncertainties are smaller.  Note
that we masked away emission from gas outside the elliptical region
shown, which appeared to have velocities that did not fit so well with
our flat axisymmetric model.  It is possible that a more sophisticated
masking algorithm, that excludes gas having anomalous velocities,
would allow our model to be extended to larger radii.

\begin{figure*}
\includegraphics[width=.9\hsize,angle=0]{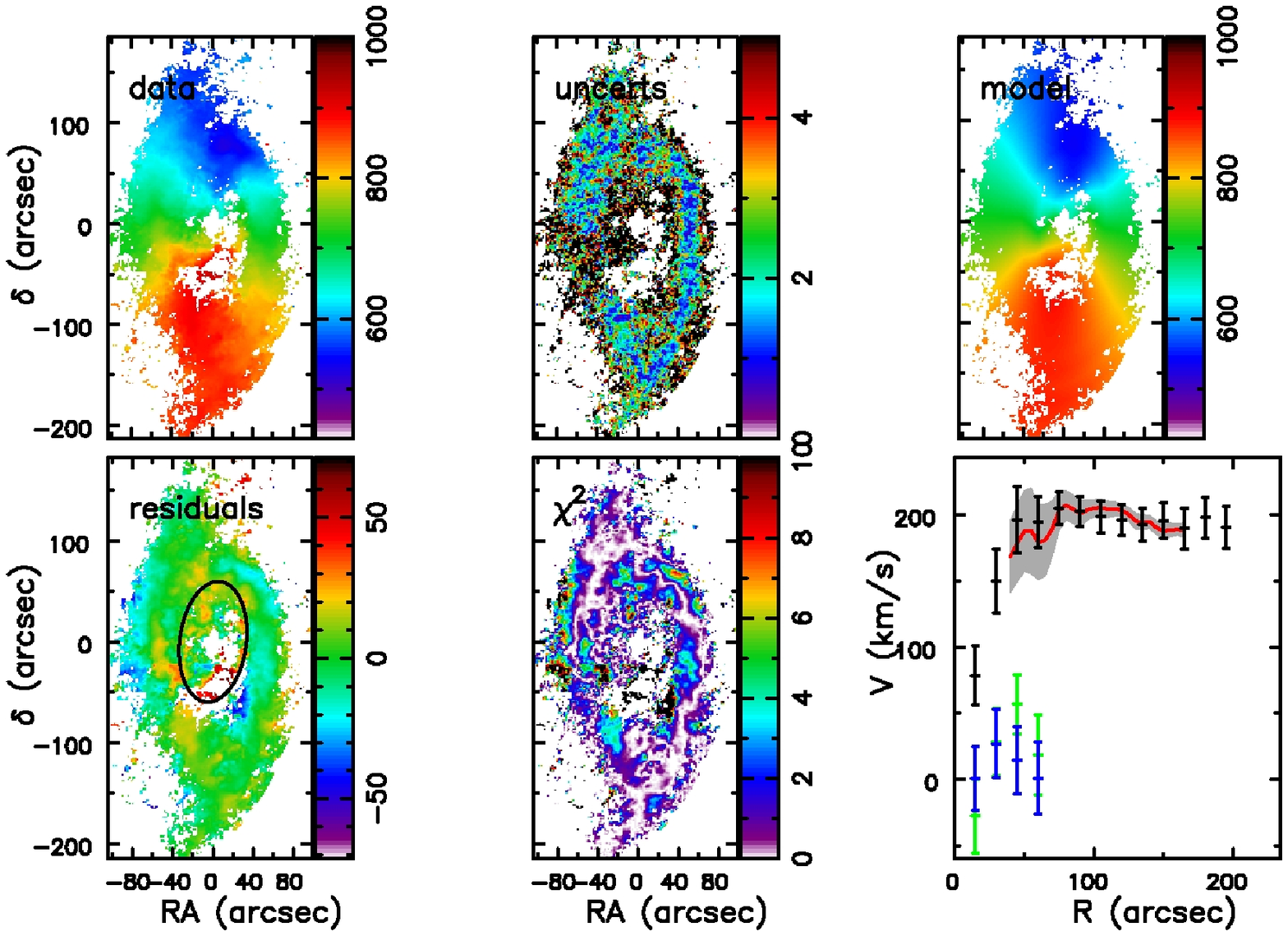}
\includegraphics[width=.8\hsize,angle=0]{cvN3627.ps}
\caption{As in Figure~\ref{fig.N2903}, but for NGC 3627.}
\label{fig.N3627}
\end{figure*}

\subsection{NGC 3627}
As this galaxy is strongly barred, we have included a bar flow in the
inner part of our fitted model, even though that part of velocity map
is rather sparsely sampled.  The resulting rotation curve is shown in
Figure~\ref{fig.N3627}, with the fitted radial and azimuthal bar flow
velocities \citep[][]{SS07} indicated, respectively, by the green and
blue points with error bars.  dB08 did not attempt to fit the barred
region, but their outer rotation curve is in reasonable agreement with
ours.

\begin{figure*}
\includegraphics[width=.9\hsize,angle=0]{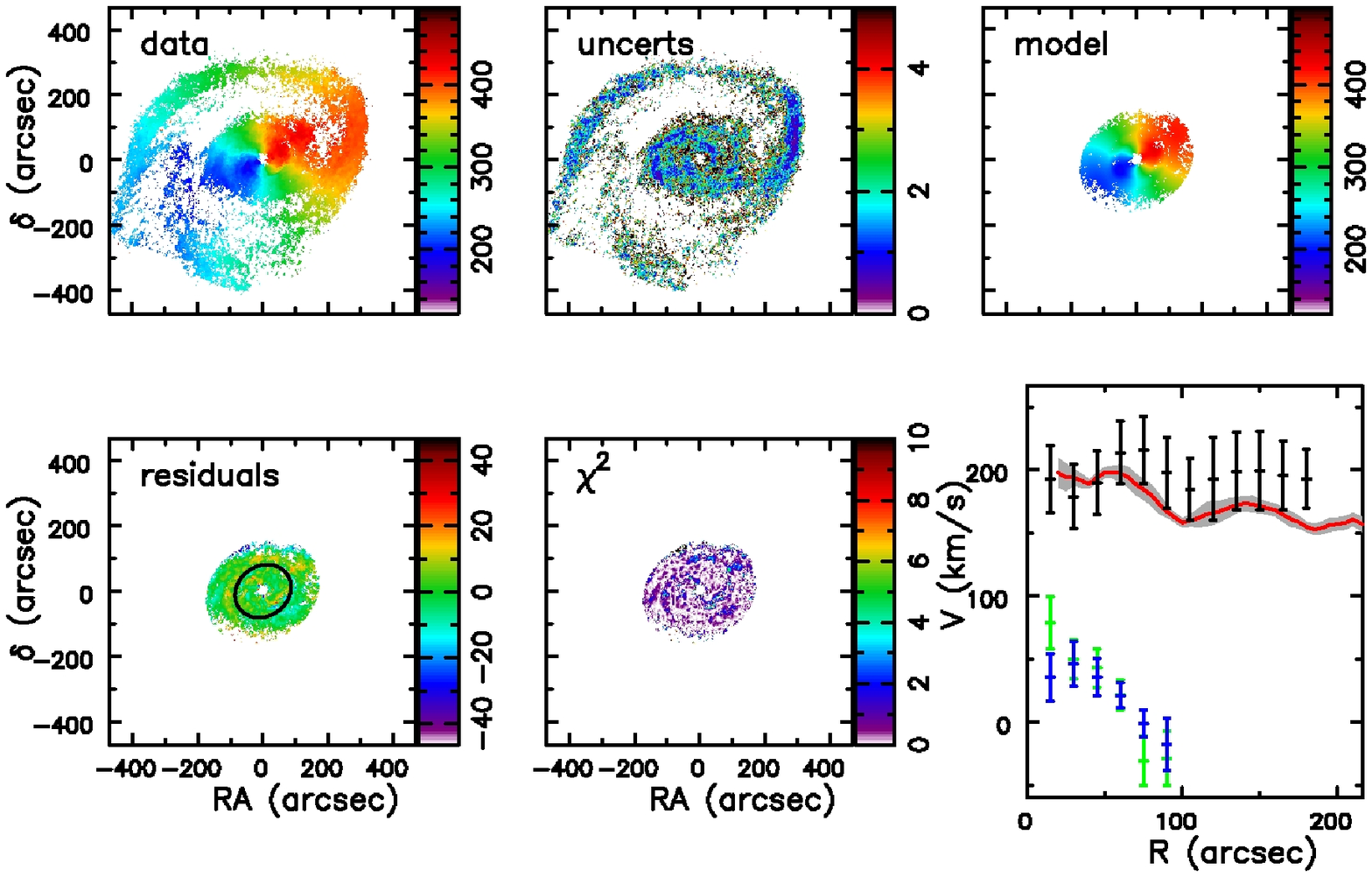}
\includegraphics[width=.8\hsize,angle=0]{cvN4736.ps}
\caption{As in Figure~\ref{fig.N2903}, but for NGC 4736.}
\label{fig.N4736}
\end{figure*}

\subsection{NGC 4736}
As the outer parts of this galaxy appear to be tidally disturbed, we
have fitted only the region inside $R = 180$\arcsec.  The resulting
rotation curve is shown in Figure~\ref{fig.N4736}, with the fitted
radial and azimuthal bar flow velocities \citep[][]{SS07} indicated,
respectively, by the green and blue points with error bars.  dB08
reported steep gradients in the apparent projection geometry within
100\arcsec, where we prefer a bar fit, and slightly more mild
variations out to $R=400$\arcsec.  Our model has generally higher
orbit speeds than theirs over the radius range we fit.  Our large
uncertainties stem from the over 5$^\circ$ uncertainty in the rather
low inclination of the inner part of this galaxy.  The covariance
plots reflect the limit we impose that the disc ellipticity cannot
fall below 0.05.

Our attempts to fit the outer parts of the map with an anti-symmetric
warp model were unsuccessful, and we have concluded that the spiral
feature encircling the north of the galaxy, together with the stream
to the SE, are tidal debris that are either not in the main plane of
the inner galaxy or not in circular motion, or both.

\begin{figure*}
\includegraphics[width=.9\hsize,angle=0]{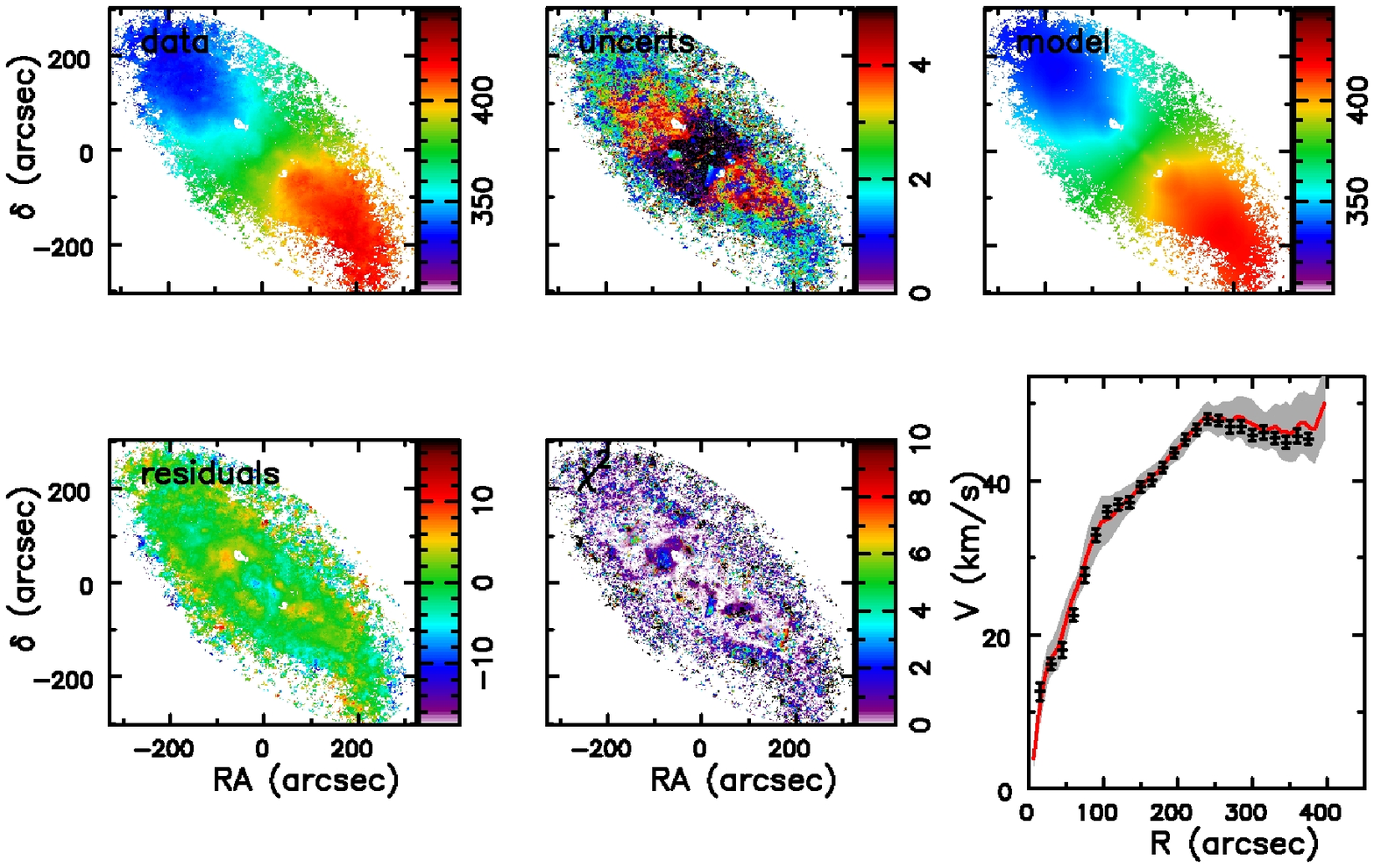}
\includegraphics[width=.68\hsize,angle=0]{cvDDO154.ps}
\caption{As in Figure~\ref{fig.N925}, but for DDO 154.}
\label{fig.DDO154}
\end{figure*}

\subsection{DDO 154}
This dwarf galaxy is extremely well fitted by a flat, axisymmetric
disc model, since we estimate uncertainties in the circular speed
(Figure~\ref{fig.DDO154}) that are substantially smaller than those of
dB08, but the rotation curves agree well.  We discuss this result
more fully in \S\ref{sec.DDO154}.

\begin{figure*}
\includegraphics[width=.8\hsize,angle=0]{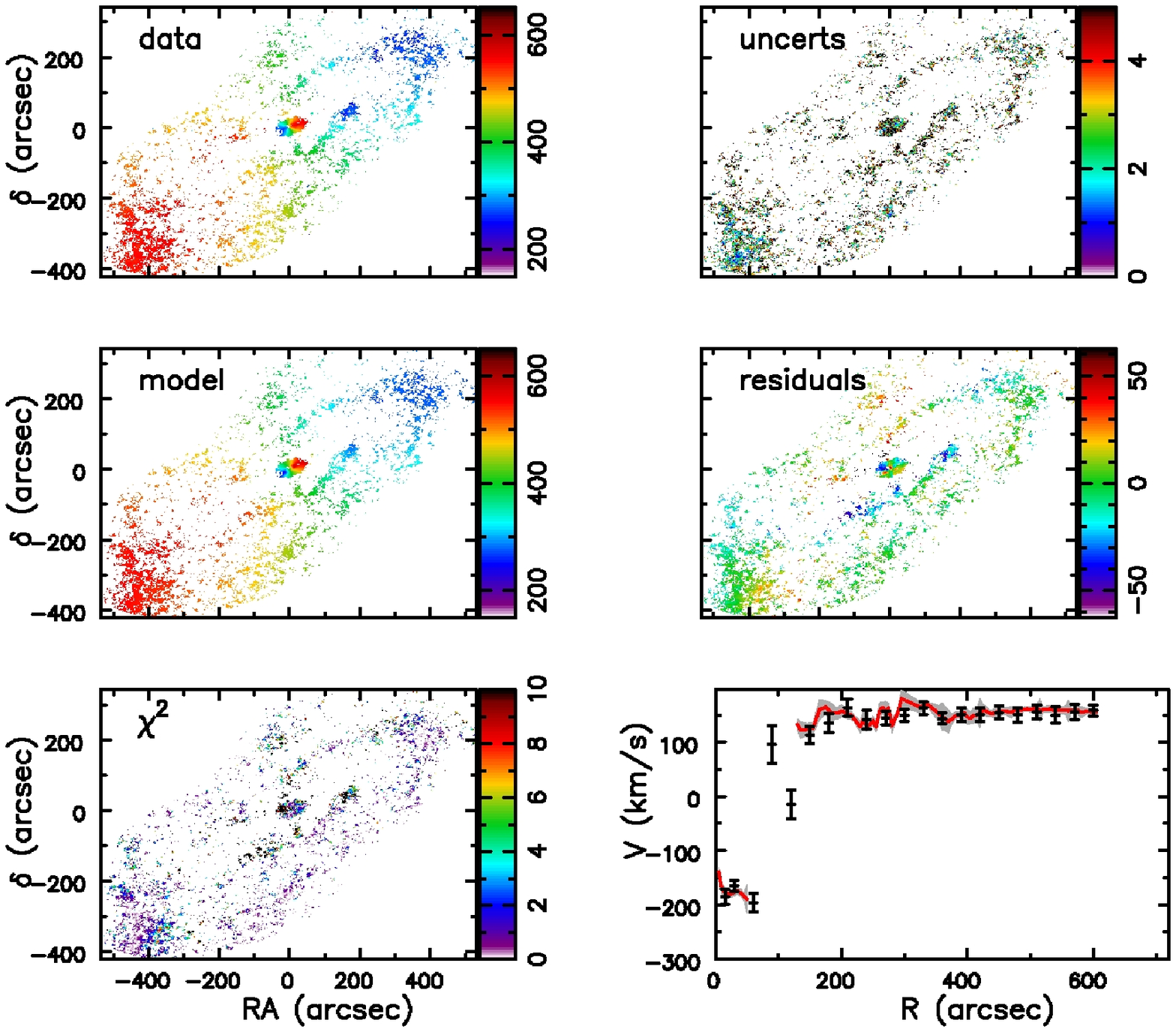}
\includegraphics[width=.68\hsize,angle=0]{cvN4826.ps}
\caption{As in Figure~\ref{fig.N925}, but for NGC 4826.}
\label{fig.N4826}
\end{figure*}

\subsection{NGC 4826}
The sparsely filled velocity map for this galaxy,
Figure~\ref{fig.N4826}, can be fitted with an axisymmetric flat disc
having a strongly counter-rotating core.  These features are in good
agreement with the fits reported by dB08, even though their fit
suggests a substantial inclination change near the centre.

\begin{figure*}
\includegraphics[width=.9\hsize,angle=0]{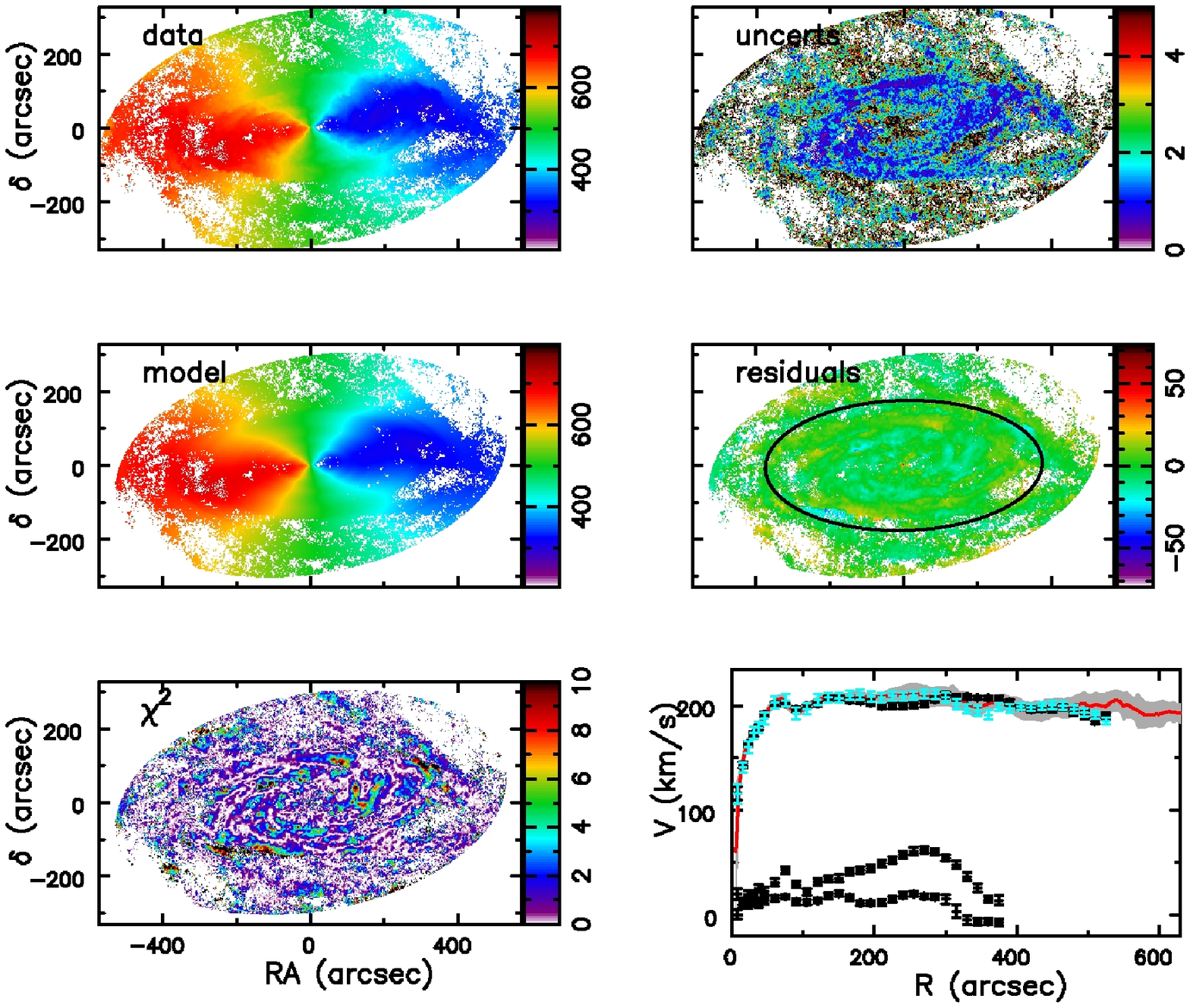}
\includegraphics[width=.7\hsize,angle=0]{cvN5055.ps}
\caption{As in Figure~\ref{fig.N2903}, but for NGC 5055.  In this
    case we have exploited the ability of \DFt\ to fit two separately
    inclined parts of the disc in one fit (see the text for a fuller
    explanation).  The small ``non-circular'' velocities of the inner
    disc returned by \DFt\ are not physically meaningful.  The cyan
    points indicate the fits to the separate inner and outer discs.}
\label{fig.N5055}
\end{figure*}

\subsection{NGC 5055}
As discussed in \S\ref{sec.results}, a flat, axisymmetric disc model
was not a good fit to the velocity map of this galaxy, and it seems
that the HI layer beyond $R_{25} \simeq 378\arcsec$ lies in a
different plane from that interior to this radius -- \ie\ the HI disc
is warped.  We have fitted the map inside and outside this radius with
two separate flat discs and the fitted circular speeds of these two
separate parts are marked in Figure~\ref{fig.N5055} by the cyan points
with error bars.  The global parameters of the fit to the inner disc
are given in Table~\ref{tab.summary} and those for the outer disc are
reported in the text.  The black points with error bars in
Figure~\ref{fig.N5055}, many of which are obscured by the cyan points,
report a fit that exploits the ability of \DFt\ to model the separate
parts of the gas layer that have differing projection geometries as an
oval flow for the inner optically bright part of the galaxy, although
we do not argue that the inner disc is intrinsically oval or that the
small ``non-circular'' velocities returned by \DFt\ are physically
meaningful.  The $8.6^\circ$ change in the fitted inclination of the
separate fits created a small discontinuity in the rotation curve
(cyan points) that is is not present in the single fit with the oval
(black points), but both sets of circular speeds are consistent within
the errors with those from dB08.  The velocity residuals are from the
single fit to both parts and are generally acceptably small.

\begin{figure*}
\includegraphics[width=.9\hsize,angle=0]{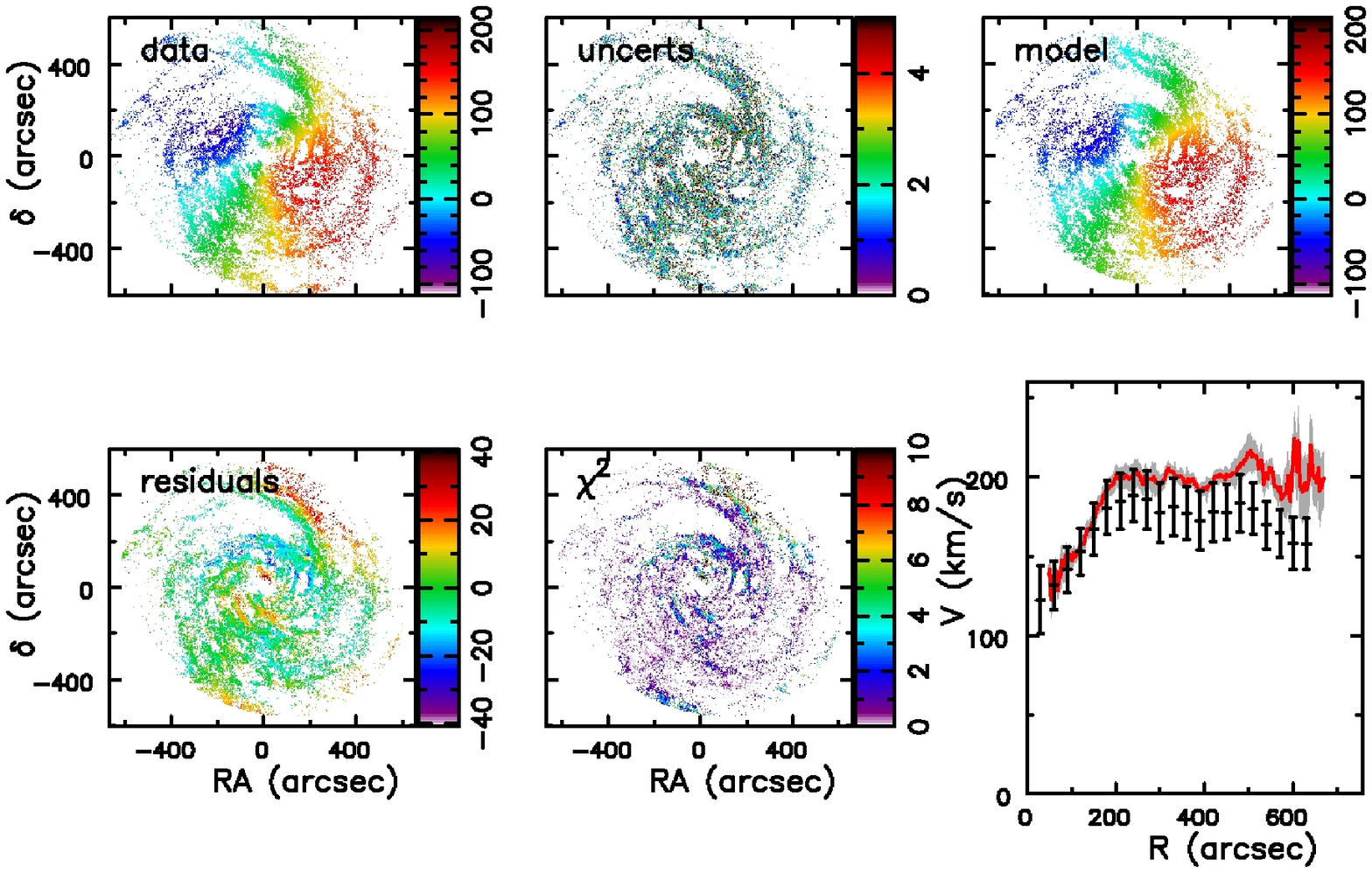}
\includegraphics[width=.68\hsize,angle=0]{cvN6946.ps}
\caption{As in Figure~\ref{fig.N925}, but for NGC 6946.}
\label{fig.N6946}
\end{figure*}

\subsection{NGC 6946}
Our fitted rotation curve for this galaxy, Figure~\ref{fig.N6946},
differs significantly from that given by dB08, especially at radii
$>200$\arcsec.  We also eliminated gas at projected radii
$>600$\arcsec, which was fitted with still lower circular speeds.  The
large uncertainties in our estimated circular speeds stem from a
significant uncertainty in the fitted inclination, for which the
1-$\sigma$ error is almost 4$^\circ$ (Table~\ref{tab.summary}), and is
also reflected in the larger coherent residuals that manifest no
tell-tale symmetry.  The flat disc assumption in \DFt\ is in tension
with the steadily varying projection parameters favoured by the
\rtc\ fits dB08.  However, our attempts to add a warp to the fitted
model made little difference, as \DFt\ preferred to apply the warp to
the last annulus only.

\begin{figure*}
\includegraphics[width=.9\hsize,angle=0]{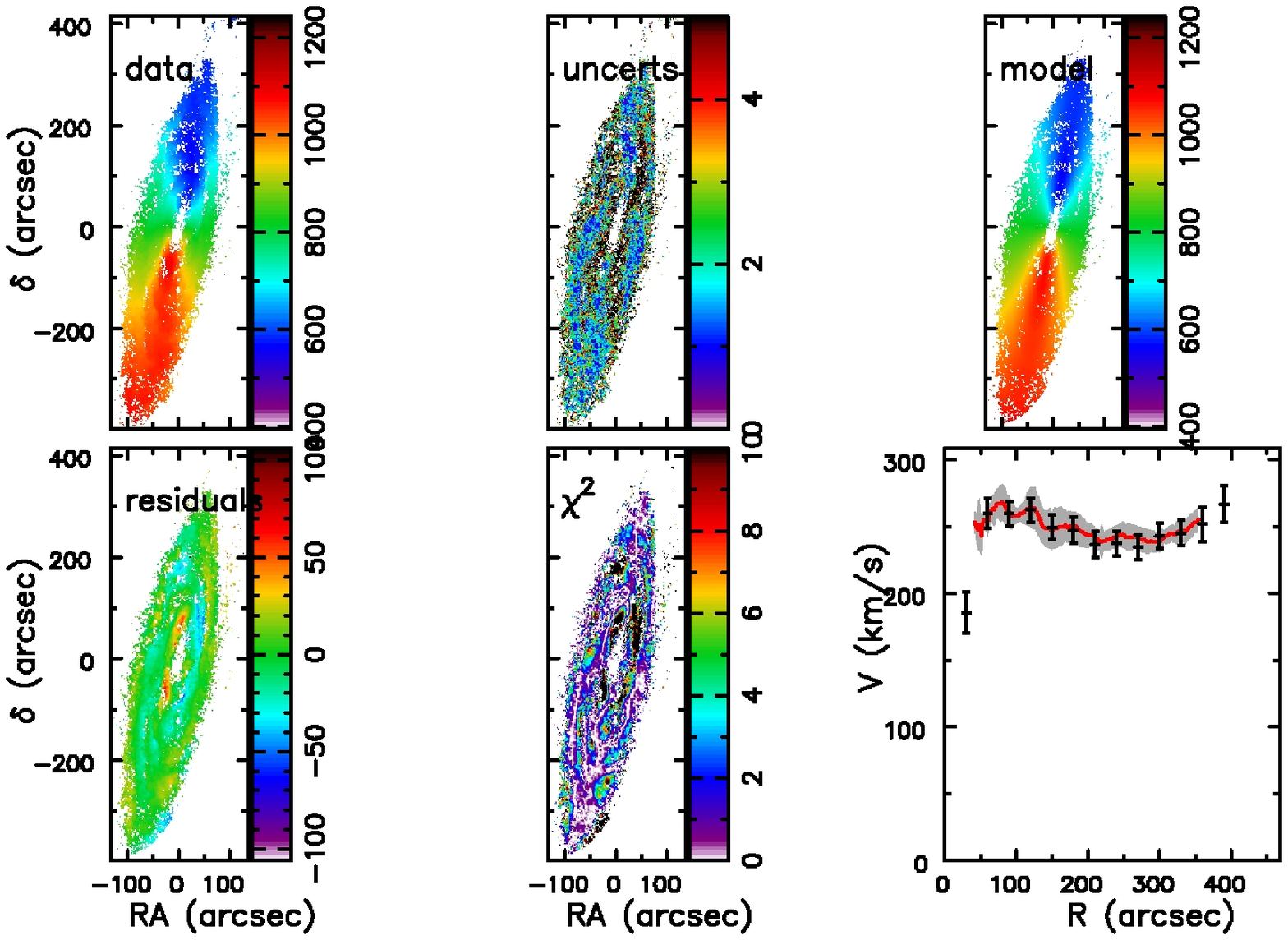}
\includegraphics[width=.68\hsize,angle=0]{cvN7331.ps}
\caption{As in Figure~\ref{fig.N925}, but for NGC 7331.}
\label{fig.N7331}
\end{figure*}

\subsection{NGC 7331}
Our fit of an axisymmetric flat disc model to this high-mass galaxy,
Figure~\ref{fig.N7331}, finds a rotation curve in good agreement with
the fit from dB08.

\begin{figure*}
\includegraphics[width=.7\hsize,angle=0]{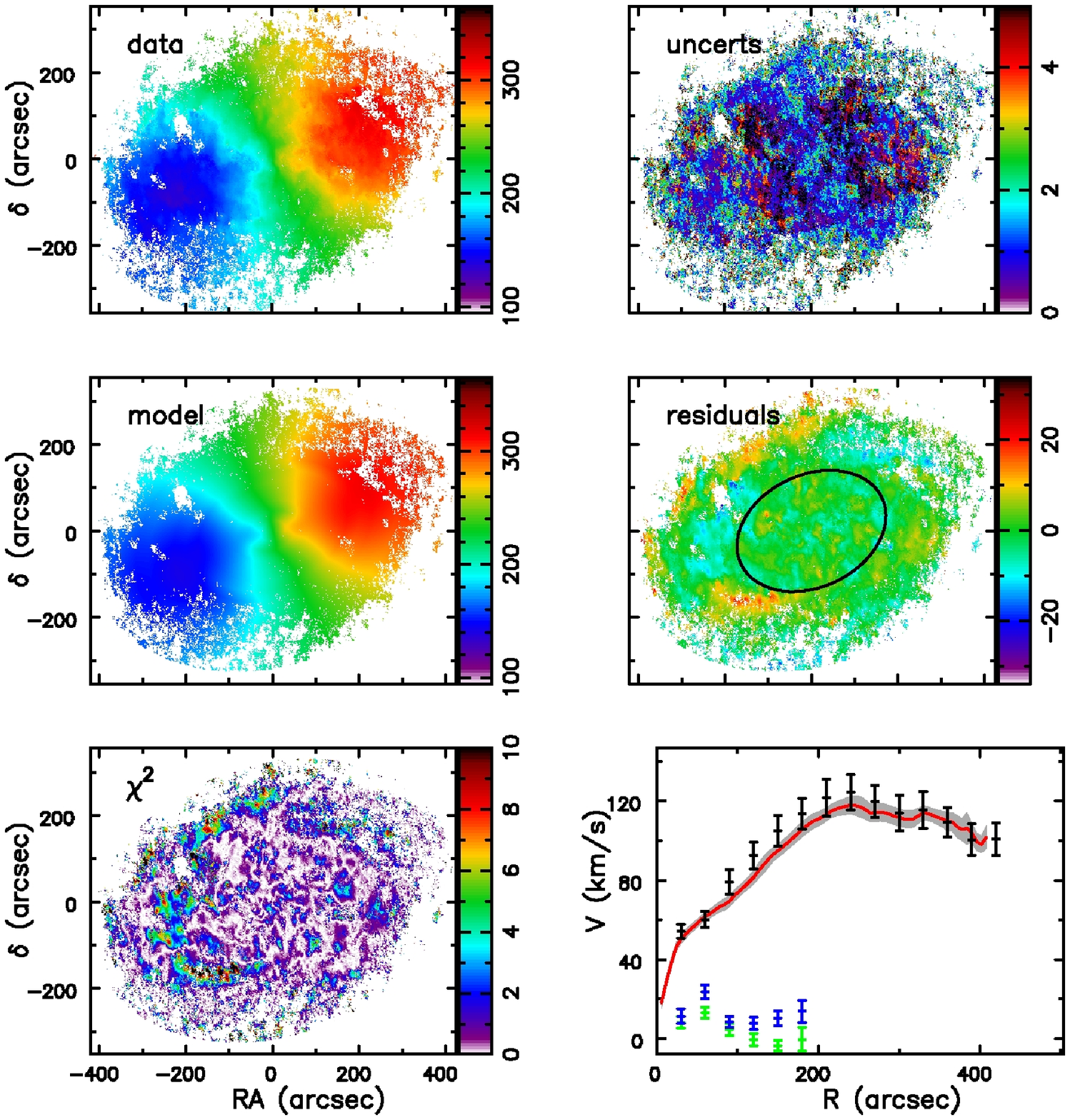}
\includegraphics[width=.8\hsize,angle=0]{cvN7793.ps}
\caption{As in Figure~\ref{fig.N2903}, but for NGC 7793.}
\label{fig.N7793}
\end{figure*}

\subsection{NGC 7793}
The velocity map for this galaxy, Figure~\ref{fig.N7793}, is better
fitted with a mild oval distortion in the inner parts, as was also
found by \citet{SZS10}.  As for NGC~2976, we find that fitting for an
inner oval in a flat disc leads to slightly higher estimates of the
circular speed over the radial range of the oval than were estimated
by dB08, who attributed the inner twisted isovelocity contours to
variations in the projection geometry.  Otherwise we confirm the
somewhat unusual overall rotation curve shape they also reported.

\bsp	
\label{lastpage}
\end{document}